\documentclass{emulateapj}

\usepackage{graphics,graphicx}

\begin{document}

\title{On the Remote Detection of Suprathermal Ions in the Solar Corona and their Role as Seeds
for Solar Energetic Particle Production}

\author{J. Martin Laming\altaffilmark1, J. Dan Moses\altaffilmark1, Yuan-Kuen Ko\altaffilmark1, Chee K. Ng\altaffilmark2, Cara E. Rakowski\altaffilmark3,
Allan J. Tylka\altaffilmark4}


\altaffiltext{1}{Space Science Division, Naval Research Laboratory, Code 7684, Washington DC 20375}
\altaffiltext{2}{College of Science, George Mason University,
Fairfax, VA 22030} \altaffiltext{3}{formerly of NRL}
\altaffiltext{4}{NASA/GSFC Code 672, Greenbelt, MD 20771}

\begin{abstract}
Forecasting large Solar Energetic Particle (SEP) events associated
with shocks driven by fast coronal mass ejections (CME) pose a major
difficulty in the field of Space Weather. Besides issues associated
with CME initiation, the SEP intensities are difficult to predict,
spanning 3 orders of magnitude at any given CME speed. Many lines of
indirect evidence point to the pre-existence of suprathermal seed
particles for injection into the acceleration process as a key
ingredient limiting the SEP intensity of a given event. This paper
outlines the observational and theoretical basis for the inference
that a suprathermal particle population is present prior to large
SEP events, explores various scenarios for generating seed particles
and their observational signatures, and explains how such
suprathermals could be detected through measuring the wings of the H
I Ly-$\alpha$ line.
\end{abstract}

\section{Overview}

An extensive body of evidence identifies shocks driven by very fast coronal mass ejections (CMEs) within a few solar radii of the Sun as the primary particle accelerators in large, gradual solar energetic particle events.
The variability of SEP events and the wide range of particle and contextual information offer
a powerful tool for exploring shock acceleration, which is ubiquitous in astrophysical plasmas.
A predictive capability for SEPs would also be of great applied interest.
Diverse lines of evidence indicate that rapid production of large intensities of high-energy particles is greatly enhanced when the pre-event environment has been primed with a population of ions having energies well above the typical thermal particle energy, usually in the range from a few to tens of keV in the solar corona. However, at present we have no direct evidence that such ``suprathermal'' ions actually exist in the corona in numbers sufficient to serve as seed particles for diffusive shock acceleration (DSA). Instead, all evidence for suprathermals in the corona is inferential.

A primary science objective must be to detect suprathermal protons
in the corona at altitudes where the production of high-energy SEPs
begins. We discuss the use of high-throughput UV spectroscopy to
quantify broadening of the Ly-$\alpha$ line resulting from charge
exchange between suprathermal protons, with energies in the
range 1 - 5 keV, and neutral hydrogen atoms in the corona. This
method provides remote-sensing of the coronal conditions that allow
large SEP events to occur. Concentrating our efforts on the seed
particle problem, we envisage a robust, simplified design optimized
to detect coronal suprathermals, with an effective area at H I
Ly-$\alpha$ being more than 500$\times$ larger than that of
instruments previously flown. Such instrumentation would also
provide a prototype for future space-weather instrumentation, and
offers great synergy with the Heliophysics System Observatory. By
determining the density of suprathermal protons in the solar corona
and the variations thereof, the connection between suprathermals and
SEP flux,  and the physical processes that create suprathermals
could be investigated directly.

In this paper we first outline the current indirect evidence for suprathermal seed particles, both observational and theoretical, in section 2. In section 3, we examine requirements on the seed
particle distribution function. We introduce the ``kappa'' distribution to model wave generation
ahead of a CME shock, and radiation transfer effects on the Lyman $\alpha$ line profile by
suprathermal atoms in the corona. The kappa distribution need not be the exact description of
the suprathermal tail, but it provides an easily calculable framework while capturing important  features of the likely seed particle distribution. Section 4 explores some of the possible mechanisms for generating seed particles, the different observational signatures they would have, and the extent to which
they can meet the requirements already discussed in section 3. In section 5, we describe some of the secondary science, concerned principally with electron acceleration, that arises. Section 6 concludes with considerations of the importance of such science to space weather and the prediction of SEP events, and the experimental implementation of such a strategy.

\section{Introduction: The SEP Effectiveness of CMEs}

At any given CME-speed, the observed SEP intensities span more than
three orders of magnitude (Reames 2000). There are many strong
observational reasons to think that an important factor in this
variability other than CME speed, is variation in the available
suprathermal seed population. The intensity of remnant energetic
particles from preceding events is the only factor discovered so far
that helps organize event-to-event variability in the peak SEP
intensities at 1 AU (Kahler et al. 1999; Kahler 2001). Gopalswamy et
al. (2004) found that large SEP events were generally preceded by
another CME erupting from the same active region within the
preceding 24 hours (see also Kahler \& Vourlidas 2005). Cliver
(2006) found that an enhanced pre-event background also favored the
occurance of Ground Level Enhancements (GLEs). The previous
CME had apparently ``pre-conditioned'' the corona to foster the
production of a large SEP event. Mewaldt et al. (2012) found that an
upper-limit on the fluence of SEP Fe at 10-- 40 MeV/nucleon can be
predicted from the Fe intensity observed at Earth at 0.01-2.0
MeV/nucleon {\it on the previous day}. Mason et al. (2005) reported
a 100-day survey at 1 AU of variation in the intensity of
suprathermal Fe ions at $\sim$30 keV/nucleon, compared to a
simultaneous survey of the bulk solar-wind Fe. Whereas the bulk
solar-wind density varies by only a factor of ten, the suprathermal
intensities span three orders of magnitude. Thus, {\it the
suprathermal variability is of the same magnitude as the scatter in
SEP intensity variations as seen by Reames (2000)}. But, of course,
these suprathermals were observed at 1 AU, not in the corona. A
number of observational studies have also identified suprathermals
from flares, which reveal themselves by distinctive compositional
signatures (i.e. high Fe/O and $^3$He/$^4$He compared to solar-wind
and coronal values), as another significant component of the
shocks's seed population in large gradual SEP events (Mason et al.
1999; Tylka et al. 2001: 2005; Desai et al. 2006ab; Tylka \& Lee
2006a; Sandroos \& Vainio 2007). In fact, evolving shock geometry,
coupled with a compound seed population comprising suprathermals
from the corona, previous CMEs, and from flares, can account for
many otherwise unexplained aspects of compositional and spectral
variability in large SEP events (Tylka \& Lee 2006ab; Sandroos \&
Vainio 2007).

\citet{tylka2006b} demonstrate that the injection profile into the heliosphere at the sun for $\sim$ GeV protons , as
deduced by \citet{bieber2004} and \citet{saiz2005} from time-intensity profiles and angular anisotropies in ground level events from the ``Spaceship Earth''
world-wide network of neutron monitors, occurs when the CME shock is within a few (2-3) solar radii. Other studies, based on satellite observations of velocity dispersion in the first arrival of particles at earth, yield times
of particle departure from the sun that agree with the neutron monitor results to within a few
minutes \citep{tylka2003,reames2009}. If we are to discover how shocks produce these very high energy particles, this is the region of the corona where we must understand not only the shocks but also the characteristics and contents of the medium through which they move.

On the theoretical side, there are also good reasons to believe that
a pre-existing suprathermal seed population can account for the
variability of CME SEP characteristics. One of the
long-standing problems in the theory of DSA is how to inject
particles from the thermal population into the acceleration process
(e.g. Malkov \& Drury 2001, Malkov \& Voelk 1995). Because DSA
involves particles being able to scatter across the shock there is a
minimum energy below which a particle cannot be accelerated by DSA.
According to Zank et al. (2006), for fast CME driven shocks
within 1 A.U. (as modeled in Zank et al. 2000), this is of order
$10^{4}$eV for parallel shocks and 20 times higher for perpendicular
shocks, but can of course vary with shock parameters. Many
authors have explored how one can achieve DSA through ``thermal
leakage'' from the thermal population (e.g. Ellison et al. 1995).
\citet{galinsky11} give a model for proton injection directly from
the upstream medium at oblique shocks, and apply it to solar wind
events. We will discuss this in more detail below, but remark for
now that such ideas may have more currency in connection to
astrophysical shocks. However, if a population of suprathermal particles already exists, as suggested by observation of elemental abundances in gradual SEP events (references cited above), then the need to energized from the solar-wind thermal population (the injection problem) becomes less relevant or non-existent.

This is demonstrated in Fig. 1, by new calculations based on work in Ng \& Reames (2008). Proton
acceleration at a parallel shock is simulated over the first 10 minutes of CME motion outward
from 3 $R_{\sun}$, for two different seed particle populations (red and blue lines on the
left hand panel). The middle and right hand panels show the ensuing proton spectrum at different
times for the lower and greater seed particle density cases respectively. The nonlinear aspect
of shock acceleration means that an order of magnitude difference in seed particle density
translates into many orders of magnitude difference in SEP densities at energies above the
Coulomb barrier; 10 - 20 MeV. Consequently, very different radiation hazards would be expected
in these two cases. For quasi-perpendicular shocks, where the injection threshold for efficient acceleration at the shock is higher (Zank et al. 2006), the need for suprathermal seed particles and the differing resulting radiation signatures would be even more acute.

\citet{giacalone05a,giacalone05b} argue that a seed particle
distribution is unnecessary, even for perpendicular shocks, in the
case that the shock is running into an irregular magnetic field.
\citet{giacalone05a} considers high Mach number shocks running into
magnetic field with fluctuations $\delta B/B$ between 0.1 and 1
using a test particle approach. \citet{giacalone05b} studies a lower
Mach number shock ($M_A=4$, closer to that appropriate for CMEs
under consideration here) using hybrid simulations, with similar
results. It might be argued that these works are simply substituting
``seed turbulence'' for ``seed particles''. \citet{gargate12}
provide a series of hybrid simulations of shocks with differing
obliquity propagating into initially undisturbed cold upstream
media. They find particle acceleration efficiency decreasing with
increasing shock obliquity, with essentially no acceleration
occurring at strictly perpendicular shocks. Observations of
suprathermal protons (47 - 65 keV) associated with shocks in the
solar wind at 1 A.U. show no significant dependence on shock
obliquity \citep{giacalone2012}, at least for the $M_A > 3$ shocks
included in the sample. Analysis of suprathermal protons observed at
a set of quasi-parallel solar wind shocks, also at 1 A.U.
\citep{neergaard2012} revealed injection energies in the range 1 - 3
keV. These authors concluded that injection proceeds directly from
the upstream Maxwellian distribution, and that a separate population
of ``seed'' particles is not necessary. However we emphasize that both these works concern
shocks observed at 1 A.U., whereas in this paper we focus on shocks much closer to the Sun.

Although theoretical opinion is far from unanimous, it is generally
recognized that the presence of a suprathermal population greatly
facilitates the effects of shock acceleration. Both observations and
theoretical studies suggest that the availability of coronal
suprathermal ions is likely to be a major factor in SEP
event-to-event variability.  But this notion must necessarily be
regarded as merely an attractive hypothesis until we find direct
evidence for coronal suprathermals, independent of the SEPs
themselves.

\section{Observational Signatures of Seed Particles}
\subsection{The ``Kappa'' Distribution}
Many authors \citep[e.g.][]{zank2006} discuss injection requirements in terms of a minimum particle
energy. Spectroscopically, what we actually observe, for instance in a line profile, is a velocity
distribution function of ions, and not an energy of a specifically suprathermal population.
Here we discuss the description of a suprathermal ion population by a ``kappa'' distribution.
Normalized over three dimensional velocity space, the function is illustrated in Figure 2 and is given by
\begin{equation}
f_{0} = {n\over 2\sqrt{2}\left(\pi\kappa\right)^{3/2}v_{Th}^3}{\Gamma\left(\kappa\right)\over\Gamma\left(\kappa -3/2\right)}{1\over [ 1 + v^2/2 \kappa v_{Th}^2]^{\kappa}},
\end{equation}
where $\Gamma\left(\kappa\right)$ is the Gamma function of argument $\kappa$.
As $\kappa \rightarrow\infty$, the distribution tends to a Maxwellian with thermal speed
$v_{Th}$. For lower values of $\kappa$, the distribution develops stronger wings; i.e. has  more particles at higher velocities than the corresponding Maxwellian. If the integration over velocity extends to infinity, we require $\kappa > 3/2$ to keep the number density finite,
and $\kappa > 5/2$
for finite energy density. Lower values of $\kappa$ require a high-velocity cutoff to keep
the suprathermal particle number and energy densities finite.

In the same way that the Maxwellian distribution can be derived from
the Boltzmann equation in terms of the coefficients of dynamical
friction and velocity diffusion, the kappa distribution is derived
with the inclusion of a velocity diffusion coefficient associated
with turbulence \citep{hasegawa85}. This is done for electrons in
lower-hybrid turbulence in Appendix B of \citet{laming2007}, and can
similarly be done for suprathermal ions. \citet{treumann08} give
more universal derivations based on an adaptation of the Gibb's
distribution (a generalized Lorentzian) appropriate for long range
correlations between particles in collisionless plasmas. Equation 1
is similar to the generalized Lorentzian distribution given by
equation 15 of \citet{cranmer1998}, and is exactly the same if we
identify $\kappa =\kappa _C+1$ and $2\kappa
v_{Th}^2= \left(\kappa _C\right)w_{\kappa _C}^2$, where $\kappa _C$ and $w_{\kappa _C}$ are
the value of kappa and the thermal width in Cranmer's expression (we have added the subscript
$C$). Thus
we are able to employ the redistribution functions for radiation
transfer given in this reference.

Many authors \citep[e.g.][]{zank2006} argue that injection energies for seed particles at quasi-perpendicular shocks should be significantly higher than is the case at quasi-parallel shocks. This suggests that ion distributions deviating more strongly from Maxwellian distributions (i.e. with a lower value of $\kappa$ giving more pronounced ``wings'' and a higher fraction of particles at higher energies) should be present at quasi-perpendicular shocks. However it is difficult to estimate what the distribution, or more quantitatively the value of $\kappa$, should be from such calculations. Here we adopt a criterion that the seed particle distribution must be capable of generating parallel propagating Alfv\'en waves when streaming upstream of a shock. This represents a minimum criterion for diffusive shock acceleration, in that upstream
particles must be able to self-generate waves, so they can be scattered back downstream.
Downstream of course, the shock itself can generate turbulence, even though the particles may also
do so.
We envisage a population of seed particles that is initially stationary in the upstream medium, being subsequently reflected further upstream upon interaction with a shock. Writing the angle between the shock velocity vector and the pre-shock magnetic field direction as $\theta _{bn}$, we evaluate
the upstream growth rate for parallel propagating waves \citep[e.g.][]{melrose1986}
\begin{eqnarray}
\nonumber\gamma &= -\int\int\int{\pi ^2q^2v_A^2\over \hbar\omega c^2}v^2\sin ^2\alpha\delta\left(\omega -\Omega -k_{\Vert}v_{\Vert}\right) \\
&\times \hbar\left({\Omega\over v_{\perp}}{\partial f\over\partial p_{\perp}}
+k_{\Vert}{\partial f\over\partial p_{\Vert}}\right)p^2dpd\left(\cos\alpha\right)d\phi
\end{eqnarray}
where $\alpha$ is the angle between the particle velocity vector and the magnetic field, and other
symbols have their usual meanings; $\omega$ is the wave frequency, $\Omega$ is the particle
cyclotron frequency, $v_A$ is the Alfv\'en speed, $v$ and $p$ are the particle velocity and momentum respectively, with subscripts $\perp$ and $\Vert$ indicating components perpendicular
or parallel to the ambient magnetic field. We restrict our attention to parallel propagating
waves, as seen in the simulations of \citet{gargate12} and the {\it in situ} observations of
\citet{bamert04}.
In a coordinate system where the magnetic field defines the
$z$ axis, the $x$ axis lies in the plane defined by the magnetic field and shock velocity vector,
and the $y$ axis is normal to this plane, the shock velocity vector has components
$\left(-v_s\sin\theta _{bn},0,v_s\cos\theta _{bn}\right)$. The streaming solar energetic particle (SEP) distribution is then
\begin{eqnarray}
\nonumber f &= f_{0}\left(1 + {3 {\bf v_s}.{\bf v} \over v^2}\right)\\
&=f_0\left(1+3{v_s\over v}\left(
\cos\alpha\cos\theta _{bn} -\sin\alpha\sin\theta _{bn}\cos\phi\right)\right).
\end{eqnarray}
Substituting this into equation 2 and integrating over $\phi$, we find
\begin{eqnarray}
\gamma & = & \int \! \int{2\pi ^3q^2 v_{A}^2 \over c^2} {v^2 \over \omega}
  {n_r\over 2\sqrt{2}\left(\pi\kappa\right)^{3/2}v_{Th}^3}{\Gamma\left(\kappa\right)\over\Gamma\left(\kappa -3/2\right)}\nonumber\\
& & \times \bigg({k_{\|}-\omega\cos\alpha /v\over p} \sin^2\alpha \cos\theta _{bn}
   {3 v_{CR}/v \over [1 + v^2/2 \kappa v_{Th}^2]^{\kappa}}  \nonumber \\
 & & {} -  {\omega \over v}\sin^2\alpha {v \over \kappa v_{Th}^2}{\kappa \over [1 + v^2/2 \kappa v_{Th}^2]^{\kappa}}
  ( 1 + {3 v_{CR} \over v}\cos\alpha\cos\theta _{bn})\bigg)\nonumber\\
  & & \times\delta (\omega -s\Omega - k_{\|}v_{\Vert})
  {p^2 dp\over m^3} d(\cos\alpha),
\end{eqnarray}
where $n_r$ is the number density of the streaming seed particle distribution. Below, $n$ will
represent the stationary background ion density.
For nonrelativistic cosmic rays, $p=mv$, $v=\Omega /(k_{\|}\vert\cos\alpha\vert) = v_{A}\Omega/(\omega\vert\cos\alpha\vert)$ and so integrating over $p$
\begin{eqnarray}
\gamma & = & \int {2\pi^3 q^2 \over mc^2} {v_{A}^4 \Omega^2\over\omega^3}
{n_r\over 2\sqrt{2}\left(\pi\kappa\right)^{3/2}v_{Th}^3}{\Gamma\left(\kappa\right)\over\Gamma\left(\kappa -3/2\right)}\times\nonumber\\
& & \bigg\{\left(3v_{CR}{\sin^2\alpha \over \vert \cos^3\alpha\vert}-3{\omega\over\Omega}v_{CR}
\cos\alpha\left|\cos\alpha\right|\right)
\cos\theta _{bn}\nonumber\\
& & \times \left[1+{\Omega^2 v_{A}^2 \over 2 \kappa \omega^2 v_{Th}^2\vert\cos^2\alpha\vert}\right]^{-\kappa} \nonumber \\
 & & {} - \left({v_{A}^3\Omega^2 \over v_{Th}^2\omega^2}{\sin^2\alpha \over \vert \cos^5\alpha\vert}
+{3v_{CR}v_{A}^2\Omega \over v_{Th}^2\omega}
{\sin^2\alpha\cos\alpha \over \vert \cos ^4\alpha\vert}\cos\theta _{bn}\right)\nonumber\\
& & \times \left[1+{\Omega^2 v_{A}^2 \over 2 \kappa \omega^2
v_{Th}^2\vert\cos ^2\alpha\vert}\right]^{-(\kappa+1)}\bigg\}d(\cos\alpha).
\end{eqnarray}
Since the region of integration is $-1\le\cos\alpha\le 1$, the terms in $\cos\alpha$ integrate to zero.
Evaluation of the remaining terms, either by analytic approximation or numerical integration, is
aided by making the substitution
\begin{equation}
{\Omega^2 v_{A}^2 \over 2 \kappa \omega^2 v_{Th}^2\vert\cos^2\alpha\vert}
= {1\over \eta^2\vert\cos^2\alpha\vert}=\tan^2x,
\end{equation}
with $\eta = \sqrt{2\kappa}\omega v_{Th}/\Omega v_A$, and using $\Omega ^2/\omega _{pi}^2 = v_A^2/c^2$ and $\beta =2v_{Th}^2/v_A^2$, the integral becomes
\begin{eqnarray}
{\gamma\over\Omega} & = & {n_r\over n}\int {\sqrt{\pi}\over 2\beta ^2}{\Omega ^4\over\omega ^4}
{\over \kappa ^2}{\Gamma\left(\kappa\right)\over\Gamma\left(\kappa -3/2\right)}
\bigg\{
3 {v_{CR}\over v_A}\cos\theta _{bn}\nonumber\\
& & \times\left(\eta^3\left|\cos^{2\kappa -3}x\right|\left|\sin x\right| -\eta {\left|\cos^{2\kappa -1}x\right|\over\left|\sin x\right|}\right) \\
&- & {v_{A}^2\Omega^2\over v_{Th}^2\omega^2}\left(\eta^5\left|\cos^{2\kappa -3}x\right|\left|\sin^3 x\right| -\eta ^3\left|\cos^{2\kappa -1}x\right|\left|\sin x\right|\right)\bigg\}dx\nonumber.
\end{eqnarray}
We evaluate equation 7 numerically, assuming $\Omega/\omega = max(M_A\cos\theta _{bn},5)$. This comes from writing $\Omega = k_{\Vert}v_s\cos\theta _{bn}=\omega M_A\cos\theta _{bn}$, and imposing a maximum value of $\omega$ as $\theta\rightarrow\pi /2$, which is discussed further below. We take the limits of integration to be $\pi/2 -\arctan\eta \le x \le \pi/2 +\arctan\eta$.
Analytically,
\begin{eqnarray}
\nonumber{\gamma\over\Omega}&\simeq {n_r\over n}{\sqrt{\pi}\over 2\beta ^2}{\Omega ^4\over\omega ^4}
{\Gamma\left(\kappa\right)\over\kappa ^2\Gamma\left(\kappa -3/2\right)}\times\\
&\bigg[3{v_{CR}\over v_A}\cos\theta _{bn}\left({\eta ^{\prime 2\kappa +1}\over\kappa\left(\kappa
-1\right)}-{\eta ^{\prime 2\kappa +3}\over\kappa +1}-{\eta ^{\prime 2\kappa +5}\over\kappa +2}-...\right)\nonumber\\
& -2\kappa\left({\eta ^{\prime 2\kappa +1}\over\kappa\left(\kappa
-1\right)}-{\eta ^{\prime 2\kappa +3}\over\kappa }  \right)\bigg]
\end{eqnarray}
where $\eta ^{\prime} = \sin\tan ^{-1}\eta <1$. In the limit $\eta <<1$
\begin{equation}
{\gamma\over\Omega}\simeq {n_r\over n}{\sqrt{\pi}\over 2\beta ^2}{\Omega ^4\over\omega ^4}
{\Gamma\left(\kappa -1\right)\over\Gamma\left(\kappa -3/2\right)}
{\eta ^{2\kappa +1}\over \kappa ^3}\left(3{v_{CR}\over v_A}\cos\theta _{bn}-2\kappa\right)
\end{equation}
suggesting $\kappa < 1.5 M_A\cos\theta _{bn}$ (to keep $\gamma >0$)
as a simple criterion for particle injection into the shock
acceleration process at oblique shocks. A stronger non-thermal tail
is required at higher shock obliquity, and at lower shock Alfv\'en
Mach number. The forgoing treatment is correct for the case of a
kappa distribution of ions flowing through a background Maxwellian,
whose particle density appears in the denominator of $v_A$. One can
easily verify that all growth and damping rates tend to zero as
$\kappa\rightarrow\infty$ so the addition of a Maxwellian has no
effect here. A streaming kappa distribution of density $n_r$ moving
through a background of density $n$ with the same kappa distribution
gives $\kappa < 1.5M_A\cos\theta _{bn}n_r/\left( n_r+n\right)$, a
more stringent requirement than the stationary Maxwellian requires.

We have tacitly assumed $n_r << n$, which need not
be the case in reality. In this case the definition of $v_A$, both in terms of the density
entering the denominator, and the frame of reference (stationary background or streaming ions)
becomes ambiguous, and requires revisiting the dispersion relations for MHD waves in the
composite plasma.

\subsection{A Case Study: The 2005 January 20 Event}
To illustrate how wave growth ahead of CME shocks might evolve, and
how the particle distributions responsible for such wave growth
might be detected, we consider the 2005 January 20 event. The flare
and associated CME and SEPs have been studied by many authors.
\citet{grechnev08} give a survey of the flare evolution from imaging
and timing data acquired by the Reuven Ramaty High Energy Solar
Spectroscopic Imager (RHESSI; Lin et al. 2002) and the
Transition Region and Coronal Explorer (TRACE; Handy et al. 1999),
and study the propagating wave that develops into a shock using
the Extreme-Ultraviolet Imaging Telescope (EIT; Delaboudini\`ere et
al. 1995) 195 \AA\ channel and white light images from the
Large-Angle Spectrometric Coronagraph (LASCO; Brueckner et al.
1995) C2 and C3. Element charge states within the CME ejecta are
studied by \citet{lepri12}, while other in situ observations are
comprehensively reviewed by \citet{foullon07}. According to
\citet{grechnev08}, the shock reached a velocity of about 2000 km
s$^{-1}$ at a heliocentric distance of 5 $R_{\sun}$, and quite
possibly achieved this speed at lower, less well observed,
altitudes. \citet{lepri12} model the behavior of the Alfv\'en speed
with altitude. The CME wave first emerges as a shock at a
heliocentric radius of about 1.3 $R_{\sun}$, where the Alfv\'en
speed is around 2000 km s$^{-1}$. The Alfv\'en speed declines
monatonically as the shock moves out, reaching Alfv\'en Mach numbers
of 2 at about 1.8 $R_{\sun}$ and 3 at 2.1 $R_{\sun}$.

\citet{grechnev08} argue that all the electromagnetic emissions they
analyze come from the flare site and, interestingly, also suggest that the flare was the location
of the associated particle acceleration. We will take the view advocated above that SEPs are
accelerated at the shock, aided by the existence of seed particles in the upstream medium, and
calculate the distribution function (i.e. the value of $\kappa$) these seed particles must
have in order to initiate particle acceleration. The 2005 January 20 event was the last in a
series of four CMEs from the same active region occurring on January 15, 17, 19, and 20,
and had by far the largest SEP productivity. \citet{grechnev08} comment that ``seed populations
from previous CMEs do not appear to be drastically different for the 17-20 January events'',
but in the absence of remotely sensed observations of the type described here in this paper it is
not clear how this could be known.

In Figure 3 we plot contours of $\gamma /\Omega$ calculated from
equation 7 for different stages of the shock wave evolution in
$\kappa - \theta _{bn}$ space, where $\kappa$ is the index of the
kappa distribution of the upstream ions, and $\theta _{bn}$ is the
shock obliquity. All panels assume $T=2\times 10^6$ K (173
eV), and a constant shock speed of 2100 km s$^{-1}$. Panel 2a gives
contours at 1.8 $R_{\sun}$, with $M_A\simeq 2$ and $v_A=1050$ km
s$^{-1}$. Panel 2b gives contours at 2.1 $R_{\sun}$ and $M_A\simeq
3$, $v_A=700$ km s$^{-1}$; 2c gives contours at 2.5 $R_{\sun}$ and
$M_A\simeq 4$, $v_A=525$ km s$^{-1}$; and 2d gives contours at 3.5
$R_{\sun}$ and $M_A\simeq 8$. The contour where the growth rate is
zero is given by the thick dashed line. Other contours are given in
units of the ion cyclotron frequency. In all cases, the requirements
on $\kappa$ are least stringent at $\theta _{bn}=0^{\circ}$, and at
the higher $M_A$ shocks with higher plasma $\beta$. At 1.8
$R_{\sun}$, we require $\kappa\simeq 3.0$ for parallel injection,
rising to 4.6 at 2.1 $R_{\sun}$, 6.2 at 2.5 $R_{\sun}$, and 12 at
3.5 $R_{\sun}$, all in approximate agreement with equation 8. Early
in their evolution, CME shocks are often expected to be
quasi-perpendicular in geometry, and at 45$^{\circ}$ we require
$\kappa\simeq 2$, 3.4, 4.5, and 9.5 respectively. In panels (c) and
(d), some of the contours show increasing $\kappa$ with increasing
$\theta _{bn}$. We have verified that this behavior arises from our
approximation concerning the value of $\Omega /\omega =
max\left(M_A\cos\theta _{bn}, 5\right)$, and that the contour where
the growth rate is zero is relatively unaffected by this. In
reality, $\omega$ should be a spectrum of waves, and not a single
value. The maximum value, 5, is motivated by the discussion of the
$M_A=3.1$ shock in \citet{gargate12}, and should probably increase
with increasing $M_A$, although it is kept constant here.

The model of \citet{galinsky11} assumes that waves generated ahead of the shock reach a steady state. This requires growth and damping rates of order $\Omega$. This can be seen to be satisfied
in Figure 3 as $\kappa\rightarrow 1.5$ out to $\theta _{bn}\sim 45^{\circ}$ \citep[the obliquity
considered by][]{galinsky11} in panels c and d, but less so where $M_A$ is smaller. This
highlights the seed particle problem. Acceleration is inferred to commence at low coronal heights,
but here shocks are generally of too low $M_A$ to self inject particles from the upstream medium.
A seed particle population is necessary at these altitudes.

\subsection{Remote Detection of Suprathermal Seed Particles}
We discuss the observation of such a suprathermal ion population directly by its effect on the profile of resonantly scattered Lyman $\alpha$ from the extended corona. Neutral hydrogen atoms remain coupled to the ionized component so long as the charge exchange rate $\int n_{i}\sigma_{cx}v d^3v < u/r$, where $n_{i}$ is the proton density, $\sigma_{cx}$, is the charge exchange cross section (e.g. Kadyrov et al. 2006), $v$ is the proton thermal velocity, while on the right hand side $u/r$ expresses the solar wind expansion velocity divided by the heliocentric radius.  This translates to $n_{i} > 10^{3}$ cm$^{-3}$, or a $r < 4 R_{\sun}$. A population of seed particles should reveal itself through extended wings on the Lyman $\alpha$ profile. Following arguments above, we
model the distribution of coronal protons and neutral hydrogen as a kappa distribution, and compute the
profile of scattered disk Ly-$\alpha$ radiation to be observed with a suitable instrument. We
follow the formalism presented in \citet{cranmer1998}, remembering the difference between his and
our definitions of $\kappa$.

We model the Ly-$\alpha$ disk profile by
\begin{eqnarray}
I\left(\nu\right)&=2.76\times 10^{17}{\Delta\nu /2\pi\over\left(\nu -\nu _0\right)^2+\Delta\nu ^2/4}\\
&\left(1-0.95\exp\left(-10^{-23}\left(\nu -\nu _0\right)^2\right)\right)
{\rm ergs~cm}^{-2}{\rm s}^{-1}
{\rm sr}^{-1}{\rm \AA }^{-1}\nonumber
\end{eqnarray}
which captures the Lorentzian wings with $\Delta\nu = 4\times 10^{11}$ Hz, and the central self-absorption.
The ion density in the extended corona is given by
\begin{equation}
n=4\times 10^8\left(R_{\sun}\over r\right)^{16}+4\times 10^6\left(R_{\sun}\over r\right)^{4.8}
+4\times 10^5\left(R_{\sun}\over r\right)^{2.5} {\rm cm}^{-3},
\end{equation}
which matches Figure 2 in \citet{cranmer1998}. The density of neutral H is based on this equation,
modified by the ratio of the radiative recombination rate \citep[from][]{seaton59} and the electron
collisional ionization rate \citep[from][]{scholz91}. The expansion velocity profile of the slow speed solar wind is taken to be
\begin{equation}
v=20 +300\log _{10}\left(r/2R_{\sun}\right) {\rm km s}^{-1},
\end{equation}
which corresponds to Figure 8 in \citet{cranmer09}.

In Figure 4, we show radiation transfer calculations following the
formalism in Cranmer (1998), for the scattering of disk Lyman
$\alpha$ radiation by hydrogen atoms in an isotropic ``Kappa''
distribution in the corona. The panels a, b, c, and d correspond to
the profiles required for injection at $\theta _{bn}=45^{\circ}$ at
1.8, 2.2, 2.5, and 3.5 $R_{\sun}$, i.e. with the required values of
$\kappa$ of 2, 3.4, 4.5, and 9.5. In each case, the thick dashed
histogram shows the kappa distribution profile, and the solid thin
histogram shows the profile resulting from resonant scattering in a
Maxwellian with temperature $2\times 10^6$ K (173 eV) for
comparison. The thick solid histogram shows the profile resulting
from having 0.1 of coronal H atoms in a kappa distribution and the
remaining 0.9 in a Maxwellian. A Maxwellian distribution (where
$\kappa\rightarrow\infty$) makes no contribution to growth or
damping rates, see equation 8. The thin dotted curve shows the disk
Ly $\alpha$ profile, to be read in
ergs~cm$^{-2}$s$^{-1}$sr$^{-1}$\AA$^{-1}$. The histograms are to be
read in counts per bin, where the bins are 0.1 \AA  wide. We have
assumed a $10^3$ s integration time with a pixel size of 1
arcmin$^2$, and an effective area varying in the range 1 - 2 cm$^2$
according to data in \citet{osantowski91}. Any instrumental
linewidth is neglected compared to the Doppler broadened coronal
linewidth. The only noise included is that due to Poisson statistics
in an assumed photon counting detector. Geocoronal absorption is not
included, but would only affect the signal in one spectral pixel
\citep{meier70}. Seed particles energized by prior shock passage
might be expected to fill the observed volume to a greater extent
than would seed particles associated with a reconnection current
sheet or outflow shock. As discussed below this last possibility is
perhaps the most plausible scenario for the generation of seed
particles, and so one should concentrate on the solid histograms
showing the 10:90 mix of seed particles and ambient plasma. Such
seed particle distributions appear to be detectable out to 2.5
$R_{\sun}$, bearing in mind that longer integration times and larger
spatial pixels are definitely feasible. Since seed particles
apparently have an effect up to 24 hrs after a prior CME (Gopalswamy
et al. 2004), integration times an order of magnitude larger are
definitely feasible, even with an observing duty cycle of 1/3 - 1/2.
Thus not only would the detection of a suprathermal population be
possible, but also its characterization in terms of density,
$\kappa$, and the likely ability to produce an SEP event when
combined with a CME shock of appropriate speed.

\section{Origins of Coronal Suprathermal Protons}
So far we have discussed the existence of seed particles from the point of view of the effects
they would have on SEP production, and not said much about how they might be produced. A number
of different ideas have been described in the literature.
For the most part, these scenarios imply different expectations in the spectral, temporal, and spatial distributions of the suprathermal protons and the Ly $\alpha$ spectra that they generate.  We briefly outline here some of these scenarios and indicate how observations could distinguish among them.

\subsection{Suprathermal Protons from Preceding CMEs}
The result of Gopalswamy et al. (2004) suggests that previous solar
activity (flare/CME) may prime the solar corona and wind with seed
particles which can then be accelerated as SEPs by a second CME.
This scenario should show significant differences in the coronal Ly
$\alpha$ profile before and after an eruptive event and would
perhaps be the easiest signature to distinguish should it be
present. The seed particles may be the result of an earlier CME, but
they are more likely SEPs associated with an earlier associated
``impulsive'' flare (Reames 1995). While the ambient solar wind
composition is simply that of the corona, exhibiting the well known
First Ionization Potential (FIP) enhancements of ions with low
first ionization potential (e.g. Feldman \& Laming 2000), impulsive
flare SEPs have a high Fe/O abundance (Tylka \& Lee 2006a)
suggesting that they are not ambient solar wind or coronal particles
energized by a shock. More probably they are accelerated via
reconnection at a flare-associated current sheet, thought to be a
plausible site for the origin of the abundance anomalies (Drake et
al. 2009). The high Fe/O particles also have high charge states
which appear difficult to achieve simply by particle stripping
during the acceleration process (Kocharov et al. 2000; Barghouty \&
Mewaldt 2000). Impulsive flare SEPs also exhibit increasing
abundance enhancement over coronal values with increasing element
mass (or possibly decreasing charge to mass). It is now known that
this anomaly continues through the periodic table to the highest
mass elements observed. Unlike the FIP effect, where an
understanding of the phenomenon is established and only details
remain to be worked out (e.g. Laming 2004, 2009, 2012), theories of
impulsive flare abundances are much less secure. It appears that
fractionation at reconnecting current sheets, either in the corona
(Drake et al. 2009) or in the chromosphere (Arge \& Mullan 1998),
may yield the increasing enhancement with ion mass.

Whether reconnection can supply particles of sufficient energies to be SEP seeds is another question. Generally, the reconnection exhaust is an outflow at the local Alfv\'en speed, which will not meet perpendicular shock injection requirements.
\citet{gordovskyy10a,gordovskyy10b} demonstrate using a test particle approach that some particles
may be accelerated to much higher energies in the electric fields of the reconnection current
sheet. They find energetic protons in power law distributions with energy proportional
to $E^{-1\rightarrow -1.5}$, corresponding to $\kappa = 2 - 2.5$. Alternatively, magnetized particles with reconnection current sheets subject to the plasmoid instability may be further Fermi accelerated by the contraction of magnetic islands (Drake et al. 2006). Although this appears more plausible for electron acceleration than for ions (e.g. Drake et al. 2009), since to participate
particles need to be initially super-Alfv\'enic, \citet{drake12} show from considering a Fokker-Planck model of the process that an $E^{-1.5}$ (corresponding to $\kappa =2.5$) spectrum
is generally obtained, at least in the limit that the acceleration time is short compared to the loss time.

Another plausible scenario appears to be ion acceleration at a fast mode shock occurring as the reconnection outflow encounters denser plasma (e.g. Selkowitz \& Blackman 2007; Mann et al. 2006),
or at fast mode shocks produced in the exhaust by an unsteady reconnection process \citep{tanuma05}. Most authors consider electron acceleration by this means, with a view to modeling hard
x-ray emissions observed during flares. We will discuss this in more detail below, but ion
acceleration should also occur here. Such a shock typically has a compression ratio $r\simeq 2$,
\citep{workman11} and is quasi-perpendicular. \citet{park12} make the point that such a shock is
typically high beta, because magnetic field has been destroyed in the upstream region by the
reconnection process and converted to heat and kinetic energy. Thus the particle injection
problem encountered for such a shock in low beta plasma (the subject of this paper for CME shocks)
does not apply for the reconnection outflow shock. Figure 6f in \citet{park12} shows the
downstream ion distribution obeying a power law $\sim E^{-2}$ up to about 20 keV. This corresponds
to $v^{-6}$ and $\kappa \simeq3$, the values to be expected for a shock with $r=2$. \citet{guo2012} find even steeper proton spectra. With
reference to Figure 2 here, the ion spectra associated with acceleration at a reconnection
outflow shock appear to be too soft to provide the required seed particle distribution, particularly at the lower $M_A$ values found closer to the sun. Particles accelerated at an outflow shock are also not expected to be confined to the reconnection current sheet. This matches with the observed onset of ion acceleration at CME shocks occurring over a large area of the shock surface.

The possibility also remains that seed particles for one CME could be produced at the forward shock of the preceding event. This would not easily produce the abundance variations with energy described above, unless flare remnant particles from the reconnection site can be injected into the CME shock; unlikely since reconnection exhausts move only at the ambient
Alfv\'en speed and ought not to be able to catch up with a shock. It is also not guaranteed to
generate hard enough suprathermal particle spectra to excite upstream waves, similarly to the reconnection outflow shock discussed above. Another theoretical argument against producing seed particles at the shock itself is that shock acceleration times are in general short compared to the shock evolution time (e.g. Zank et al. 2006, Ng \& Reames 2008). Given the diffusion coefficients, $\kappa_{xx}$, of Zank et al. the acceleration timescale,
$\tau_{acc}\propto \kappa_{xx}/v_{shock}^2$ is of order minutes to hours for a perpendicular shock. This is ample time to produce SEPs from a single shock.
The fundamental obstacle to shock acceleration appears to be the injection of particles. If a shock were easily able to do that, the survey of Gopalswamy et al. (2004) would have turned out very differently.

\subsection{Suprathermal Protons from Ubiquitous Reconnection Processes}
In the modeling of Tylka \& Lee (2006), small impulsive SEP events, with their distinctive abundance distortions, were used to model the ``flare'' component of the suprathermal seed population.  But impulsive events are not necessarily associated with a visible X-ray flare.  Instead, they are produced by reconnection processes in the corona, which may or may not be sufficiently energetic to accelerate electrons and generate X-rays.  It is therefore possible that such reconnection activity goes on in the corona at nearly continuous level, generating suprathermals but at levels that are far below detectability at interplanetary spacecraft.  In this case, we would expect to see ubiquitous and omnipresent evidence for suprathermal protons, perhaps modulated in intensity by the general level of flare activity. \citet{masson13}
consider the release of solar-flare accelerated particles onto open solar wind field lines as
a flux rope erupts and undergoes interchange reconnection with neighboring field lines.


\subsection{Suprathermal Protons from the Decay of Flare-Generated Neutrons}
Feldman et al. (2010) reported a surprisingly large fluence of solar neutrons observed by the Messenger spacecraft at 0.48 AU following an M2 flare.  Solar neutrons are produced when energetic particles accelerated in a flare loop travel downward and impact the chromosphere, with the resulting nuclear collisions liberating neutrons, some of which travel outward from the Sun.   Feldman et al. also suggested that lower-energy neutrons that decay into protons while traversing the corona might provide a pool of suprathermal seed protons for shocks.  Although there are concerns about background-contamination in the Messenger measurements (Share et al. 2011), the notion of seed protons from the decay of flare-generated neutrons may nevertheless be viable.  If this process were to occur, we should expect the enhanced suprathermal proton population to appear within a few hours of the flare.  The spatial distribution would probably be symmetric about the
normal to the flare location, with a neutron energy spectrum (and subsequent decay-proton energy spectra) dependent on the angle to the normal.  If gamma-ray observations (from RHESSI or Fermi) were also available, modeling should be able to account for the observed radial distribution and energy spectrum of the suprathermal protons (Murphy et al. 2007, 2012).

\subsection{Magnetic Pumping Model}
One of the most intriguing discoveries in recent years has been the
ubiquitous presence of suprathermal tails on the momentum
distributions of ions in the solar wind, with a common spectral
shape of $p^{-5}$ in momentum. The common spectral shape occurs in
the quiet solar wind far from shocks, in disturbed conditions
downstream from shocks, and throughout the heliosheath. In the Fisk
\& Gloeckler model (2007, 2008; see also Fisk, Gloeckler \&
Schwadron 2010) this suprathermal tail is accelerated from the core
distributions in compressive turbulence that is thermally isolated.
It is further suggested that this
mechanism may also serve to generate suprathermal ions in the
corona, thereby providing seed particles for shock acceleration.  If
this model is correct, ubiquitous and omnipresent wings on the Ly
$\alpha$ line profile should be observable, although the magnitude
may vary with local turbulence levels.  In addition, the $p^{-5}$
spectrum will correspond to a distinctive spectral shape in the Ly
$\alpha$ wings. It is also sufficiently hard to provide seed
particles able to generate waves upstream of shocks, out to $\theta
_{bn}\simeq 30^{\circ}$ for $M_A=2$ and $\theta _{bn}\simeq
55^{\circ}$ for $M_A=3$.

\subsection{Observing Strategy}
Gopalswamy et al. (2001) and Mann et al. (2003) model the Alfv\'en speed above an active region by assuming the active region magnetic field to be a dipole superimposed on the large scale solar magnetic field. Depending on the CME speed, a shock is expected to form in the region 1.2 -- 3 $R_{S}$. Above the first critical Alfv\'en Mach number (Edmiston \& Kennel 1984), a number of order 1.5 -- 2 depending on shock geometry, the shock must accelerate particles. Slit positions chosen below and above the height where many propagating disturbances associated with CMEs are expected to steepen into a shock and begin to accelerate particles would allow discrimination between different production scenarios. Taking these slit positions at 1.5 and 3.0 $R_{\sun}$, if the CME shock produces seed particles we expect to see a signal first in the 3.0 $R_{\sun}$ slit. If the reconnection current sheet is responsible, the 1.5 $R_{\sun}$ slit should show the clearest signal. The calculations in Figure 4 are based on an assumed $10^3$ s integration in an arcmin$^2$ pixel,
which corresponds to the time taken for a 1000 km s$^{-1}$ CME to travel 1.5 $R_{\sun}$. One could clearly use shorter integration times during a CME event to improve the discrimination of the timing of the appearance of Ly $\alpha$ wings. Figure 5 replots some of the examples from
Figure 4 at shorter integration times, 10 s and 100s. Signatures of seed particles are still
visible at these shorter cadences.

Such timing discrimination would also be important for inferring the presence of seed particles
due to flare-generated neutrons. The other two scenarios, suprathermals from ubiquitous reconnection events and the $p^{-5}$ proton spectrum resulting from thermally isolated compressive
turbulence have no specific time dependence, but would have distinctive spectral signatures to
observe. The observing strategy would probably involve long integration times during quiet
times to establish the possible existence of a seed particle population and to characterize its
properties (i.e. value of kappa and number density). Given this information,
one could predict the CME shock velocity required in order for an SEP hazard to develop, using the
analysis presented above in section 3.1. During CME events observations should switch to shorter
integration times to take advantage of the relative timing of the appearance of
suprathermal ions (if any) in the regions of the corona viewed by the different spectrometer slits.

We end this
section by mentioning another requirement: seed particles produced by a prior CME must remain
with sufficient number density in the low corona to affect the evolution of a second CME for
a period of up to about 24 hours, a timescale about an order of magnitude longer than that
expected if seed particles were swept out with the solar wind. Here we estimate the
diffusion coefficient required to restrict the outward motion of energetic particles.

The relevant diffusion coefficient is \citep[e.g;][]{blandford87,rakowski2008}
\begin{equation}
D = {p^2c^2v\over 3\pi q^2 U\left(k_{\Vert}=\Omega /v_{\Vert}\right)} = {16c^2v^2\Omega\over
3\omega _{pi}^2\delta v_{k_{\Vert}}^2},
\end{equation}
where $\Omega = qB/mc$ is the particle cyclotron frequency, $U\left(k_{\Vert}=\Omega /v_{\Vert}\right) = \rho\delta v_{k_{\Vert}}^2/2$ is the wave energy density at parallel wavenumber $k_{\Vert}=\Omega /v_{\Vert}$, and other symbols have their usual meanings. Taking $B=1$G,
a number density of $n=10^6$ protons cm$^{-3}$, and $\left<v_{\Vert}\right>\sim v/2$,
$D\sim 3\times 10^{13} v^2/\delta v_{k_{\Vert}}^2$ cm$^2$s$^{-1}$. Putting the diffusion timescale
$t=l^2/D$ with lengthscale $l\sim 1 R_{\sun}$, we find $t\simeq 2\times 10^8 \delta v_{k_{\Vert}}^2/v^2$ s. Setting $t\simeq 8\times 10^4$ s, $\delta v_{k_{\Vert}}/v \simeq 2\times 10^{-2}$. So trapping a 10 keV proton ($v\sim 10^8$ cm s$^{-1}$) requires $\delta v_{k_{\Vert}}\simeq 20$
km s$^{-1}$. This is of similar order of magnitude, but smaller than the thermal line broadening at coronal temperatures, and so probably not unambiguously detectable as a signal of seed particles.

\section{Suprathermal Electrons}
Many of the mechanisms discussed above also accelerate electrons. While the necessity of electron
seeds for further shock acceleration is controversial, the observation and characterization of any
suprathermal electron distribution may further aid in understanding the processes responsible
for any suprathermal ion population.

Recent observations (e.g. Krucker et al. 2010) have indicated that the energy in accelerated electrons during a flare can be similar to that in the magnetic field, and that the whole population of electrons is energized. This argues strongly in favor of electron acceleration by a Fermi process in contracting magnetic islands in a reconnecting current sheet(Drake et al. 2006, where such a degree of electron energization was actually predicted) rather than acceleration at an outflow shock (where only a small fraction of the particles would be expected to be accelerated).
In the limit of strong magnetic field, or electron plasma $\beta\rightarrow 0$, a $v^{-5}$
electron distribution is predicted, becoming a steeper power law when the back-reaction of the
electron pressure on the acceleration process is included. Consequently reconnection in lower
electron plasma $\beta$ conditions is expected to put more of its released magnetic energy into
accelerated electrons.

In Figure 6 we show various illustrations of how heated electrons
distort the extended wings of the Lyman $\alpha$ profile through
Thomson scattering. Panels a and b show the profiles of a Maxwellian
at temperature $2\times 10^6$ K (173 eV), and kappa
distributions with $\kappa = 2.5$ and 5, intended to match the range
of electron spectra produced by the simulations in
\citet{drake2006}, and observed by \citet{oka12}. Panel a gives the
case at 1.8 $R_{\sun}$, and panel b at 2.2 $R_{\sun}$. It can be
seen that $\kappa =5$ in Thomson scattering is indistinguishable
from a Maxwellian. The $\kappa = 2.4$ profile however is distinct,
and potentially detectable. We have assumed a $10^5$ s integration
time, with emission concentrated in a 0.1 $\times 1$ arcmin$^2$
pixel, assuming that we would be observing electrons heated in a
reconnection current sheet of about 0.1 arcmin width. The histograms
have a 1 \AA\ bin size. Panels c and d show similar calculations for
the same locations, but with Maxwellian electrons with temperatures
of $2\times 10^6$, $4\times 10^6$, and $8\times 10^6$ K. These
different profiles are more distinguishable from each other, and
offer further diagnostics of reconnection processes.

\section{Summary}

SEPs are a major radiation hazard for both robotic and manned missions.  The episodic nature of the SEP hazard presents engineering challenges different from those of the other ionizing-radiation populations found in space.  Discoveries about the nature, variability, and origin of coronal suprathermal protons would be significant breakthroughs in understanding the provenance of SEPs. We have argued based on prior observational evidence and theoretical considerations that the presence
of a suprathermal seed particle population is an important factor, and maybe the crucial factor,
in determining whether a CME becomes ``SEP effective''. The observation of such a population,
through for example the line profile of Lyman $\alpha$ emitted from the extended corona, would
be a major step forward in our understanding.
We further anticipate that similar instrumentation, ultimately deployed as a monitoring mission, would greatly enhance operational forecasts of SEP events, and also provide notice of ``all-clear'' periods, through the presence or absence of seed particles as revealed by spectroscopy of the H I Lyman $\alpha$ line profile.

Given that Cycle 24 is presently predicted to have a sunspot maximum in June-July 2013 that will be smaller than any other observed during the space age, it is difficult to extrapolate from previous experience to say how many SEP events would be observed in the next few years.  However, it is worth noting that some of the largest SEP events in the historical record, including August 1972, October 2003, January 2005, and December 2006, occurred in the declining phase of the cycle.
An experiment launched in the next few years would be expected to be on-orbit while the new Solar Orbiter and Solar Probe missions are operating.  These missions, combined with others likely to be operational in an extended-mission mode (SDO, Hinode) will provide data on the solar and inner-heliospheric context of such observations.

Some of our work also has astrophysical applications. \citet{raymond10} observe a kappa distribution in emission in H $\alpha$ from the forward shock wave in Tycho's supernova remnant.
This arises as neutral H in the upstream medium passed through the shock, and is excited to emit
H $\alpha$ before it is ionized postshock, and thus retains its preshock distribution function.
\citet{raymond10} use a different definition of $\kappa$, so their quoted value of $\kappa\simeq 2$ corresponds to $\kappa\simeq 3$ in our work. At
this value, we estimate a fraction of order $10^{-3} - 10^{-2}$ of the shocked plasma particles are
able to join a seed particle population. Raymond et al. (2010) express some concern that
such a value of $\kappa$ implies very efficient particle acceleration. Considerations of seed particle
injection outlined in this paper suggest that this need not be the case, especially at oblique shocks close to the perpendicular limit, though of course the Alfv\'en Mach number is likely much
higher.

\acknowledgements This work was supported by basic research funds of the Office of Naval Research.
JML and CER also acknowledge support under grant NNH10A009I from the NASA Astrophysics Data Analysis Program. We also thank Ron Murphy for reading a draft of the paper and providing many helpful
comments.

\begin{figure}
\centerline{\includegraphics[scale=0.8]{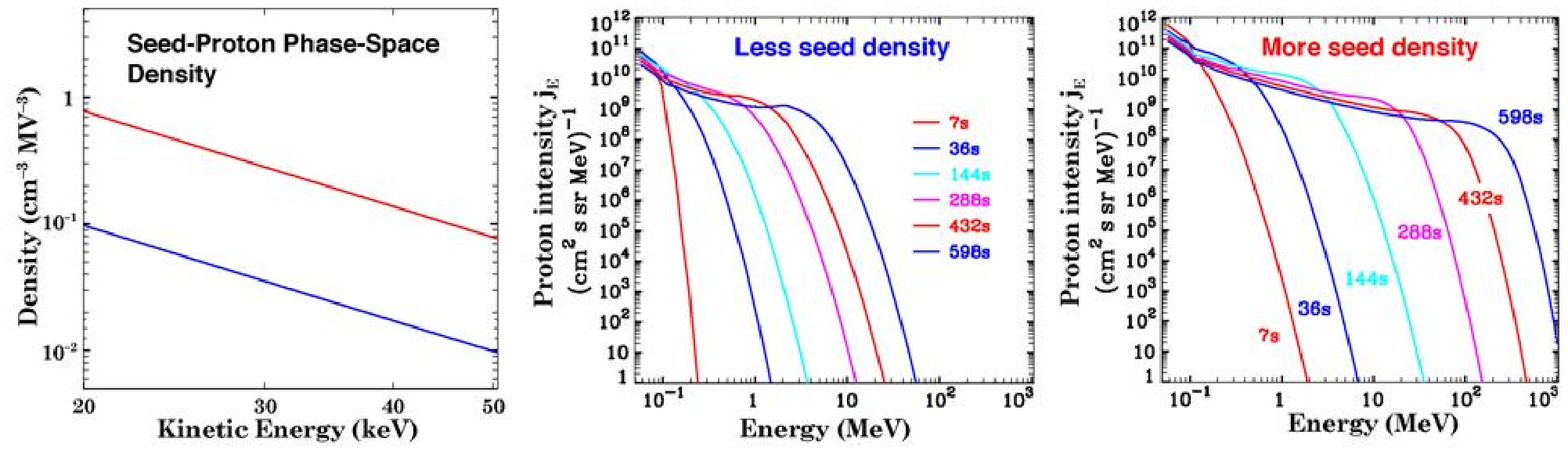}}
\caption{Evolution over the first ten minutes of the proton spectrum at the shock (based on Ng \& Reames 2008) as a CME moves outward from $\sim 3 R_{\sun}$ at 2000 km s$^{-1}$ for two different levels of input suprathermal seed protons (first panel).  Different colors in the second and third panels correspond to different elapsed times in seconds since the start of particle acceleration,  as given in the labels. These calculations take into account growth of proton-amplified Alfv\'en waves, which is a non-linear process.  Accordingly, the simulation results are in real units and are not adjusted by any arbitrary normalization factors.
\label{fig1}}
\end{figure}

\begin{figure}
\centerline{\includegraphics[scale=1.0]{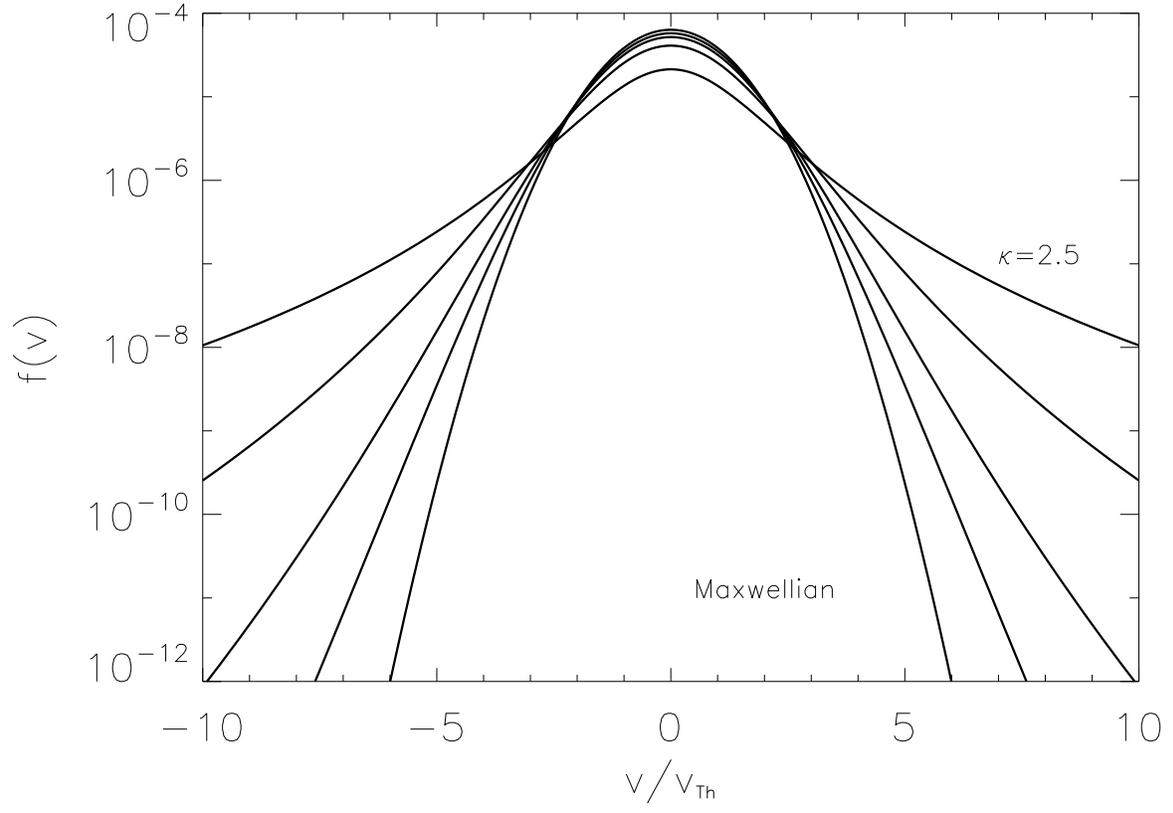}} \caption{The
``kappa'' distribution for various values of the $\kappa$ parameter.
Curves are shown for $\kappa = 2.5$, 5, 10, 20, and also the
limiting case $\kappa \rightarrow\infty$ where a Maxwellian is
obtained. The velocity axis is given in units of the thermal speed
$v_{Th}$ entering equation 1. \label{fig2}}
\end{figure}

\begin{figure}
\centerline{\includegraphics[scale=0.5]{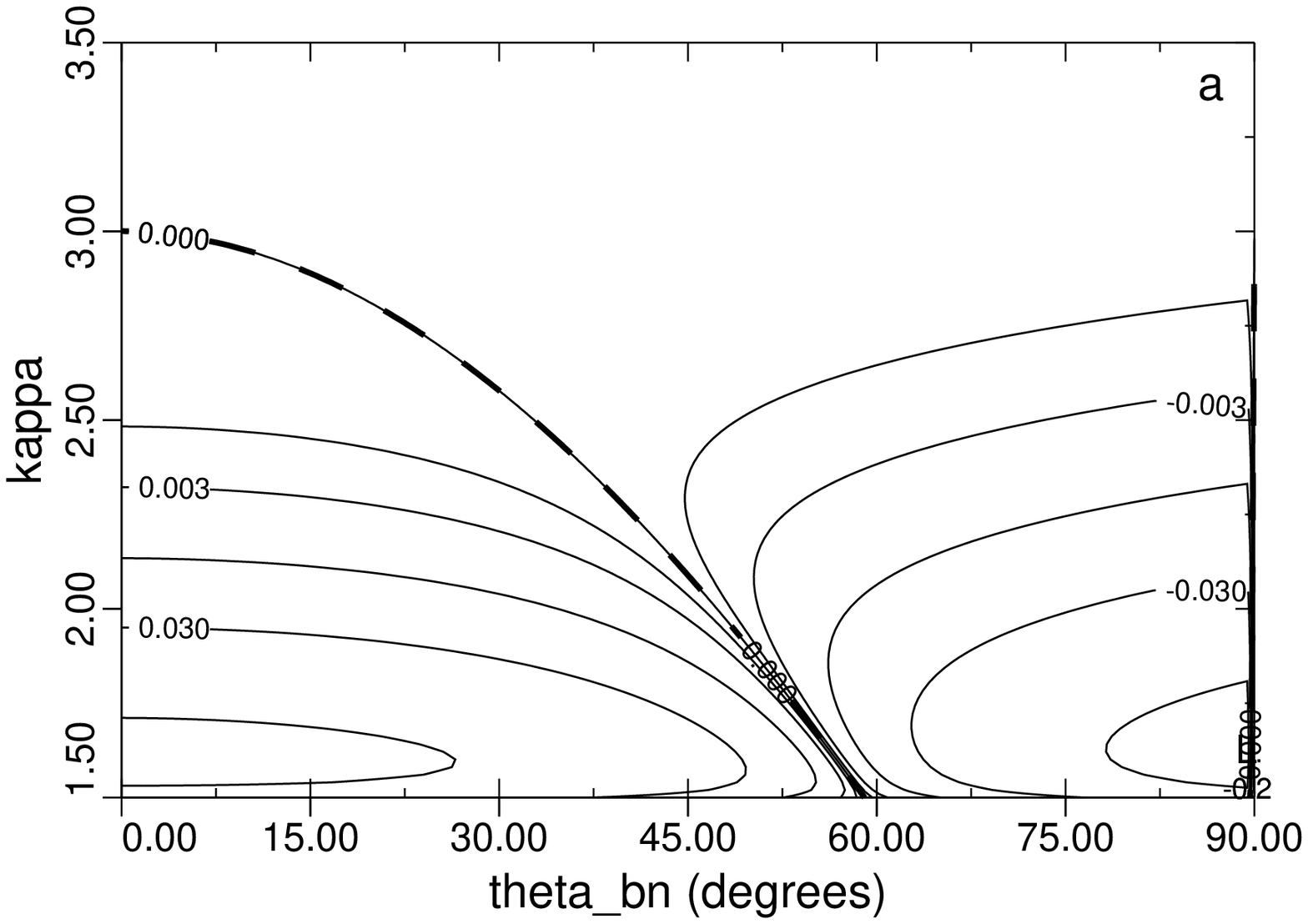}\includegraphics[scale=0.5]{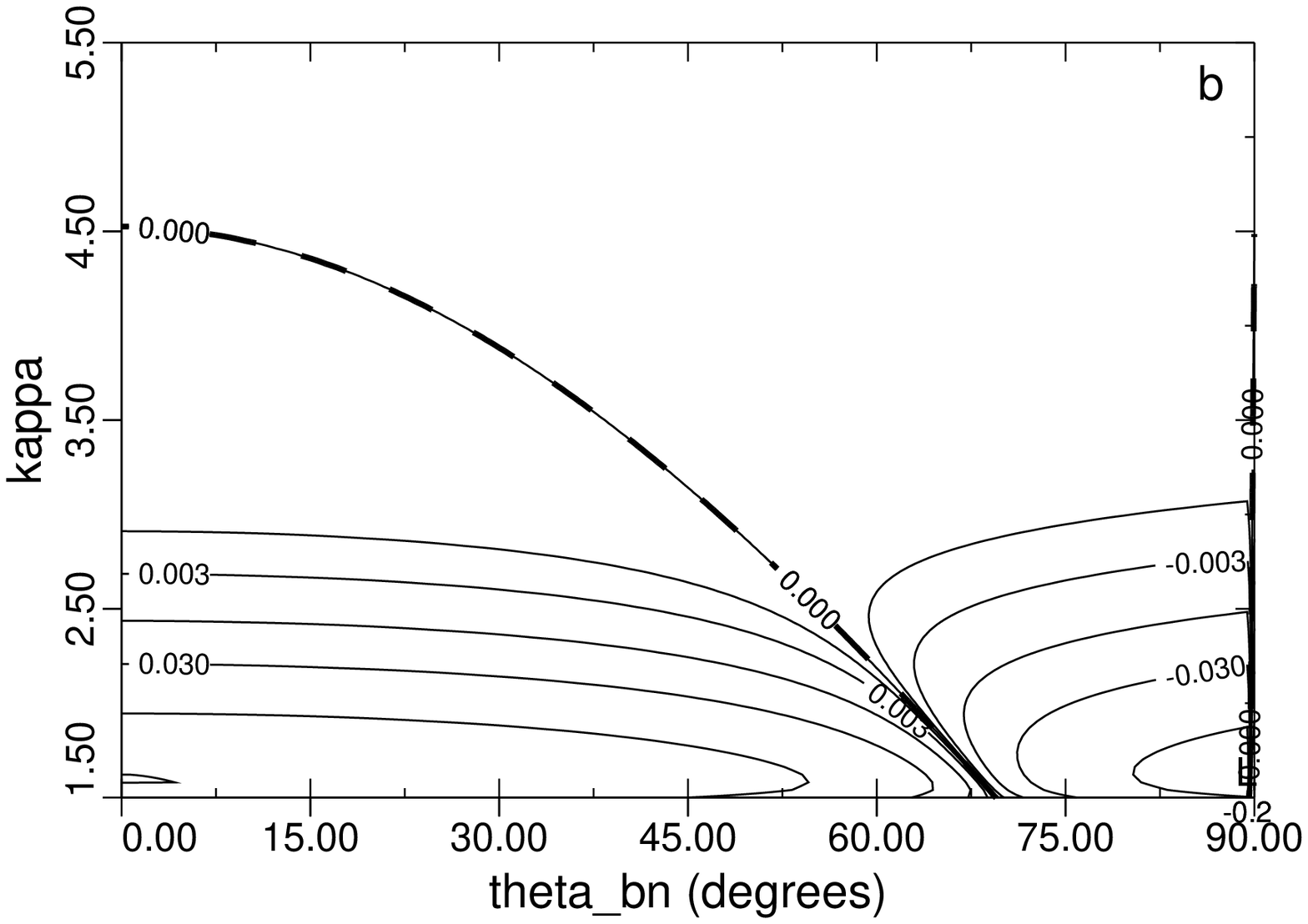}}
\vskip 20pt
\centerline{\includegraphics[scale=0.5]{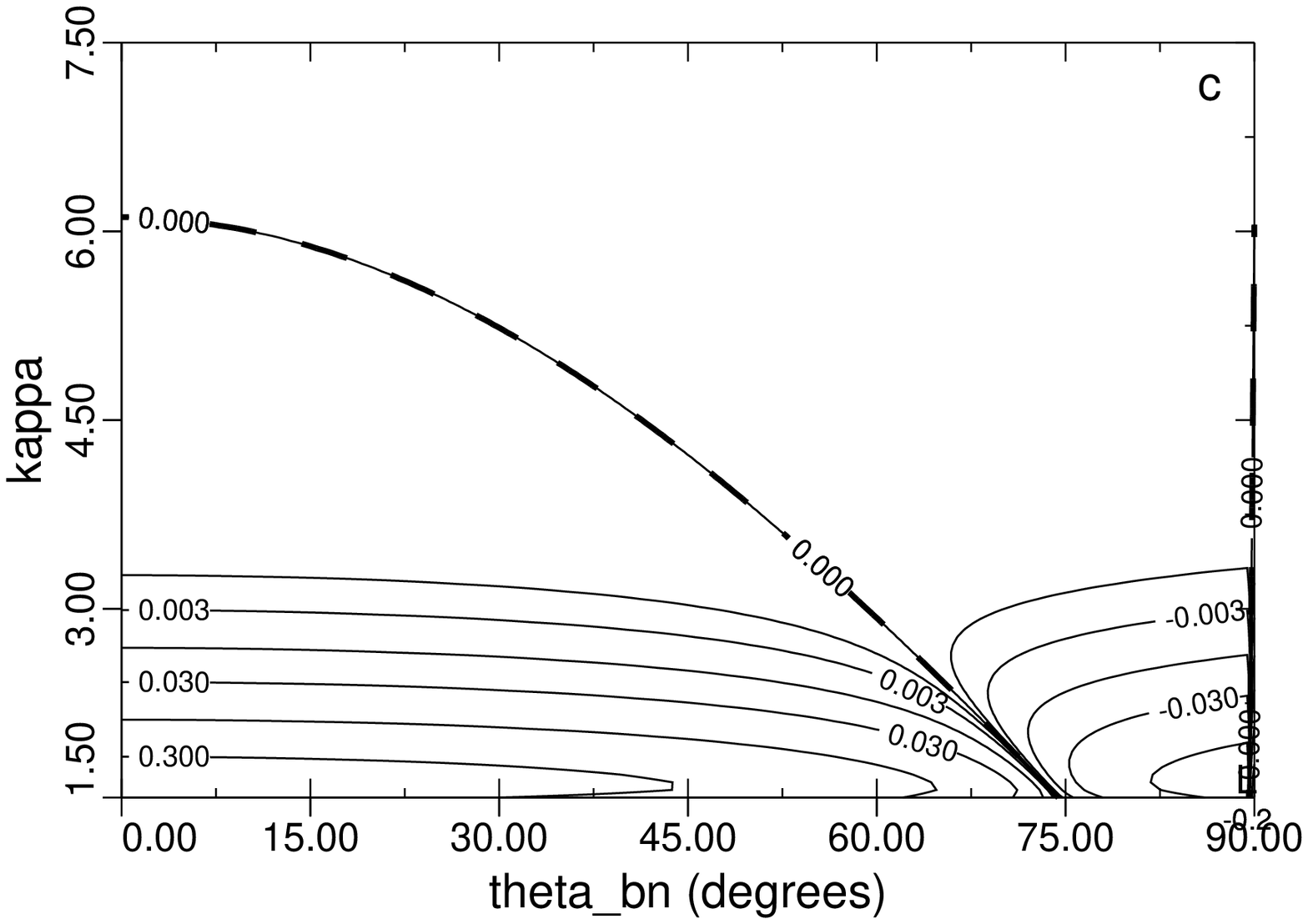}\includegraphics[scale=0.5]{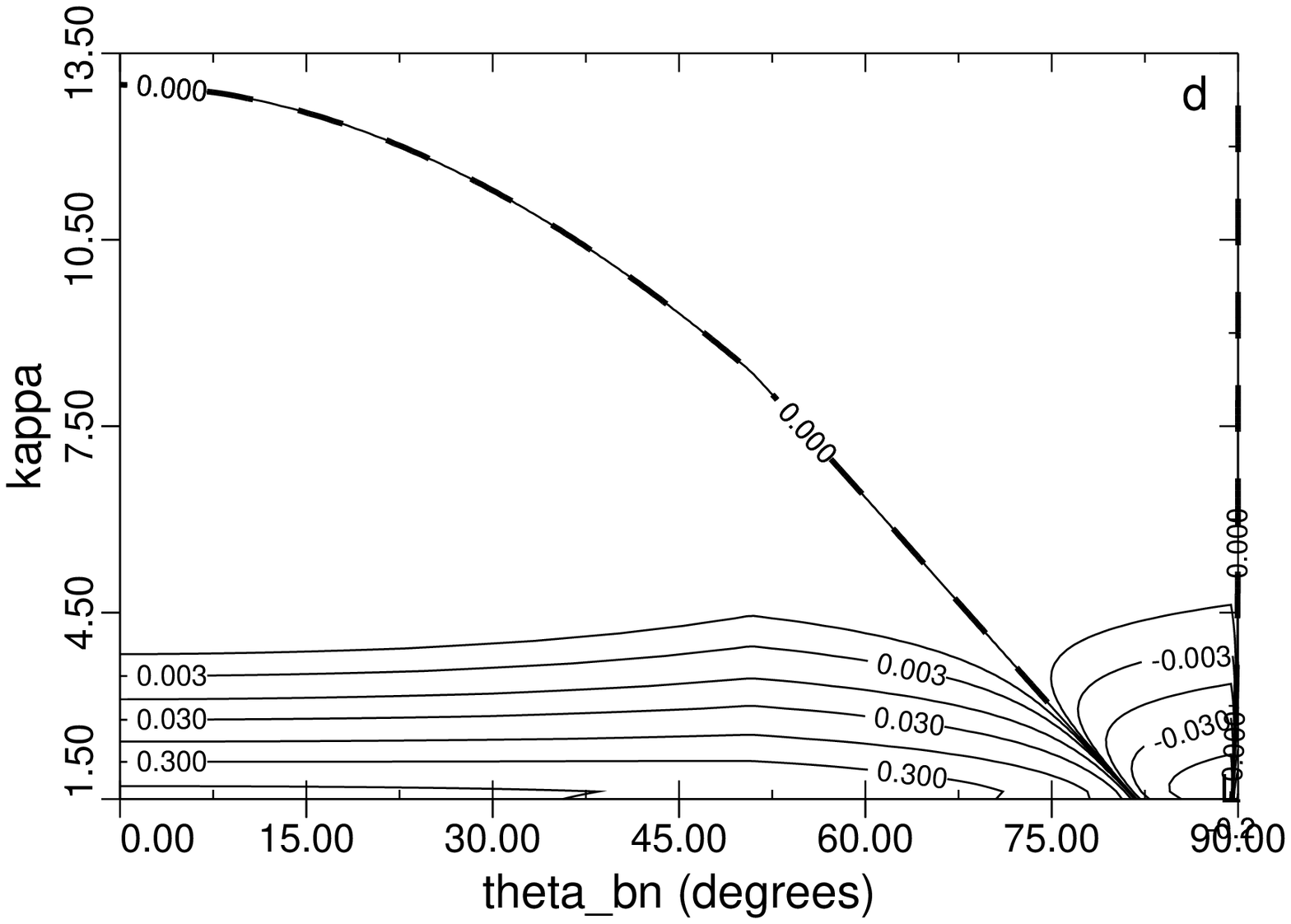}}
\caption{Contours of upstream wave growth rate given by equation 7
in $\kappa - \theta _{bn}$ space, where $\kappa$ is the index of the
kappa distribution of the upstream ions, and $\theta _{bn}$ is the
shock obliquity. All panels assume $T=2\times 10^6$ K (173
eV), and give in order the cases $M_A=2$, $V_A=1050$ km s$^{-1}$;
$M_A=3$, $V_A=700$ km s$^{-1}$; $M_A=4$, $V_A=525$ km s$^{-1}$;
$M_A=8$, $V_A=262.5$ km s$^{-1}$; corresponding to parameters for
the shock of the 2005 January 20 event at 1.8, 2.1, 2.5 and 3.5
$R_{\sun}$ respectively.
The contour where the growth rate is zero is given by the thick dashed line. Other contours are given in units of the ion cyclotron frequency. In all cases, the
requirements on $\kappa$ are least stringent at $\theta _{bn}=0^{\circ}$, and at the higher $M_A$
shocks with higher plasma $\beta$. At 1.8 $R_{\sun}$, we require $\kappa\simeq 3.0$ for parallel
injection, rising to 4.6 at 2.1 $R_{\sun}$, 6.2 at 2.5 $R_{\sun}$, and 12 at 3.5 $R_{\sun}$. At these
altitudes, injection can extend out to about 35$^{\circ}$, 60$^{\circ}$, 65$^{\circ}$, and
80$^{\circ}$ respectively.
\label{fig3}}
\end{figure}

\begin{figure}
\centerline{\includegraphics[scale=0.5]{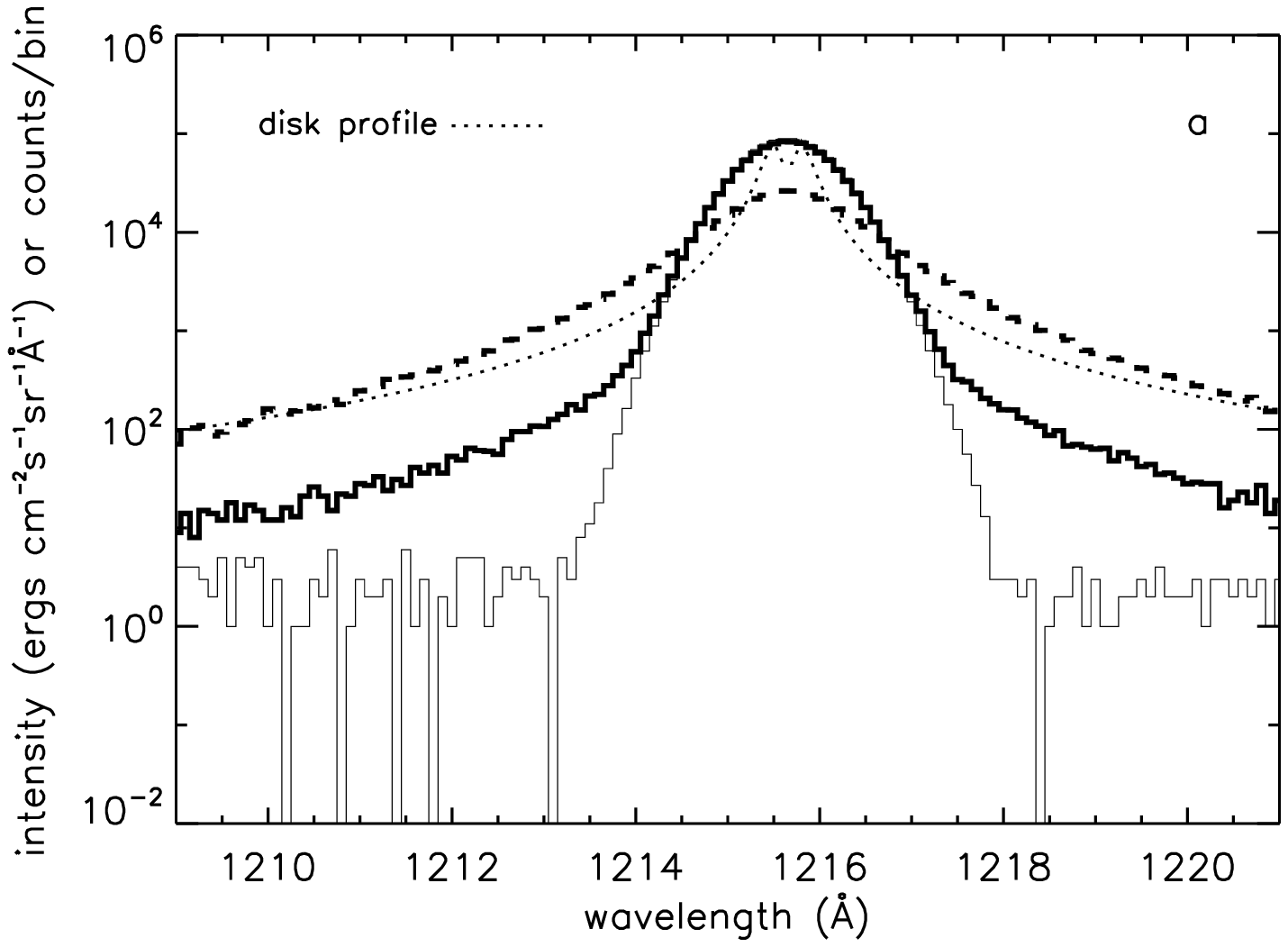}\includegraphics[scale=0.5]{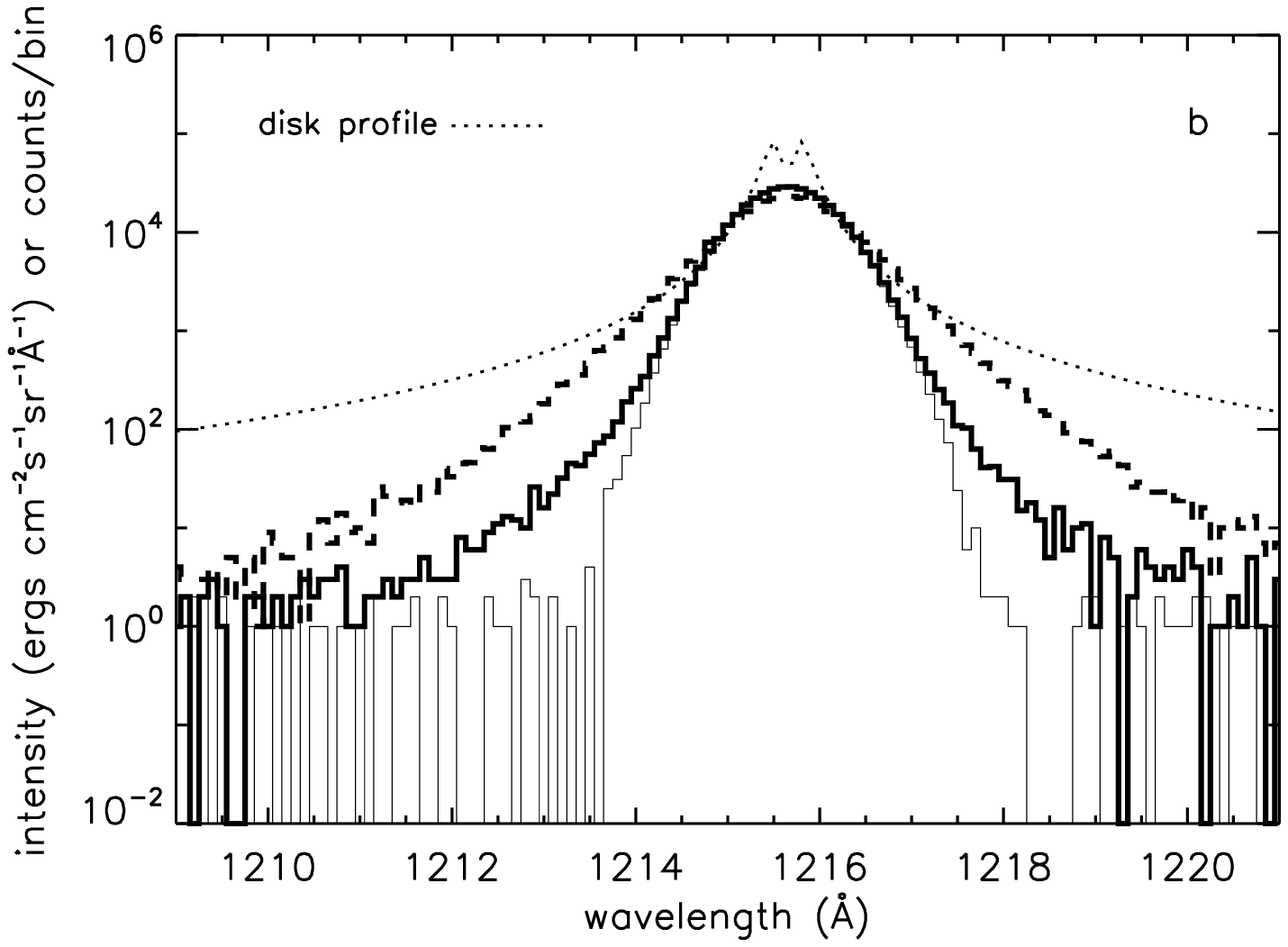}}
\vskip 20pt
\centerline{\includegraphics[scale=0.5]{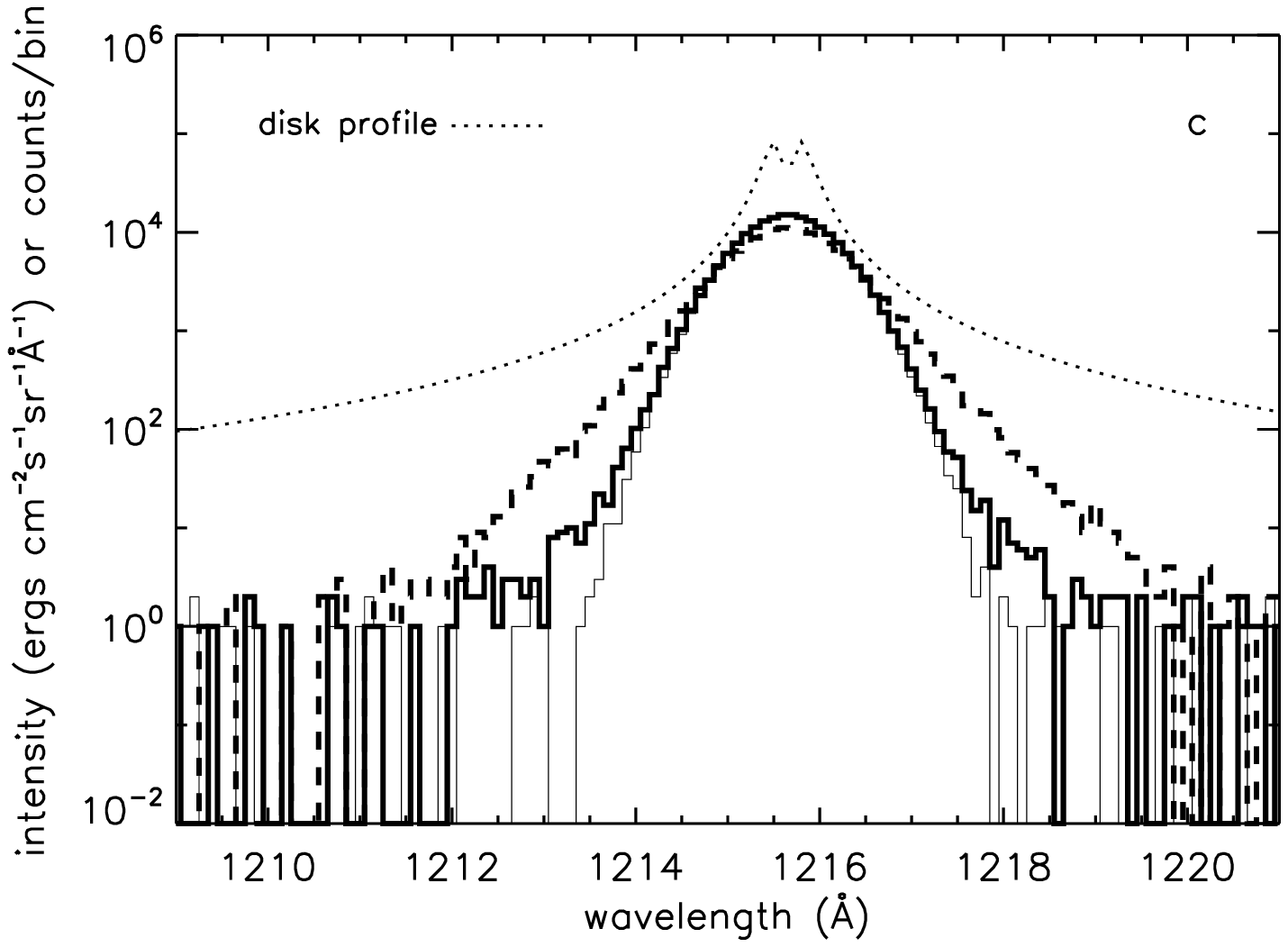}\includegraphics[scale=0.5]{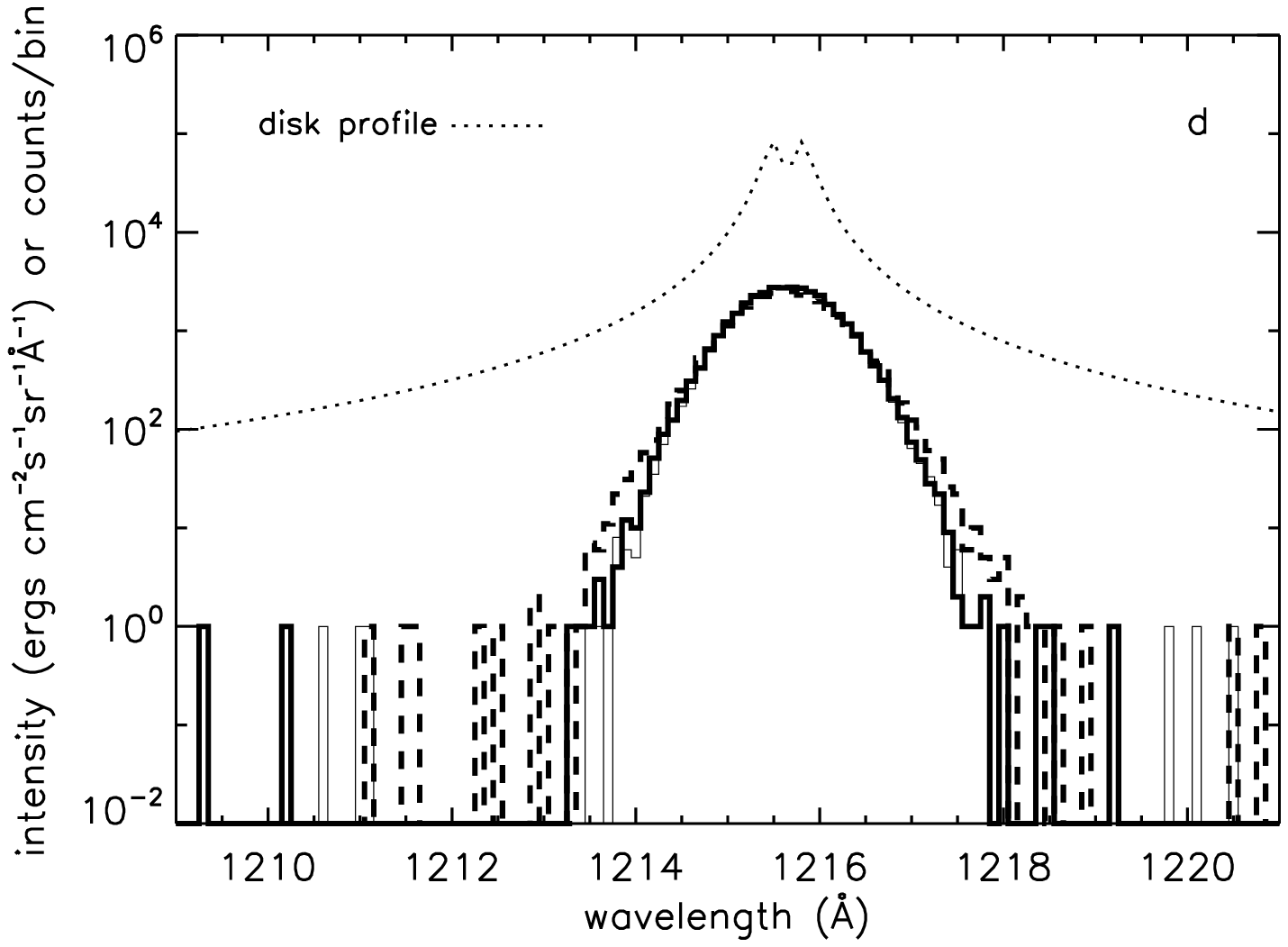}}
\caption{Simulated Lyman-$\alpha$ profiles corresponding to the
$\kappa$-distributions  required for injection into the parallel
shock regions in the panels of Figure 3. The radiant profile is
indicated as a dotted line, to be read on the intensity axis in
units ergs cm$^{-2}$s$^{-1}$sr$^{-1}$\AA $^{-1}$. The dashed
histogram shows the result of scattering of this radiation in a
$\kappa$-distribution of hydrogen atoms with value 2, 3.4, 4.5, and
9.5 respectively. The thin histogram shows the result in each case
for a Maxwellian, with the thicker histogram showing an intermediate
result with 90\% of the hydrogen atoms along the line of sight in a
Maxwellian distribution, and 10\% in the $\kappa$-distribution. We
have assumed Poisson statistics in a photon counting detector, with
effective area of $\sim 1$ cm$^2$, and $10^3$ s integration time.
All histograms give the counts in 0.1 \AA\ bins, in a spatial pixel
of 1 arcmin$^2$. Such seed particle distributions are easily
detectable out to 2.5 $R_{\sun}$, bearing in mind that longer
integration times and larger spatial pixels are definitely feasible.
\label{fig4}}
\end{figure}

\begin{figure}
\centerline{\includegraphics[scale=0.5]{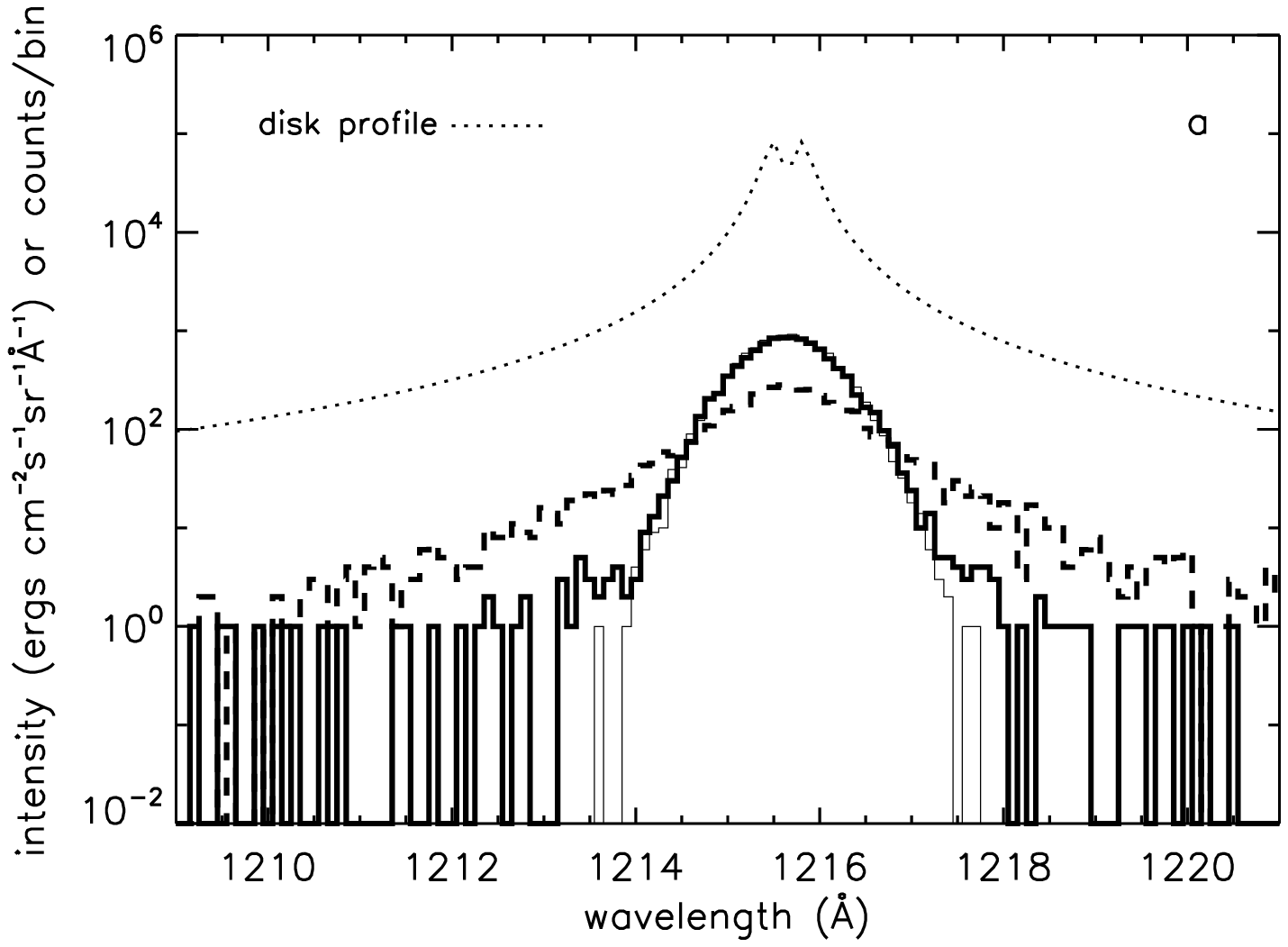}\includegraphics[scale=0.5]{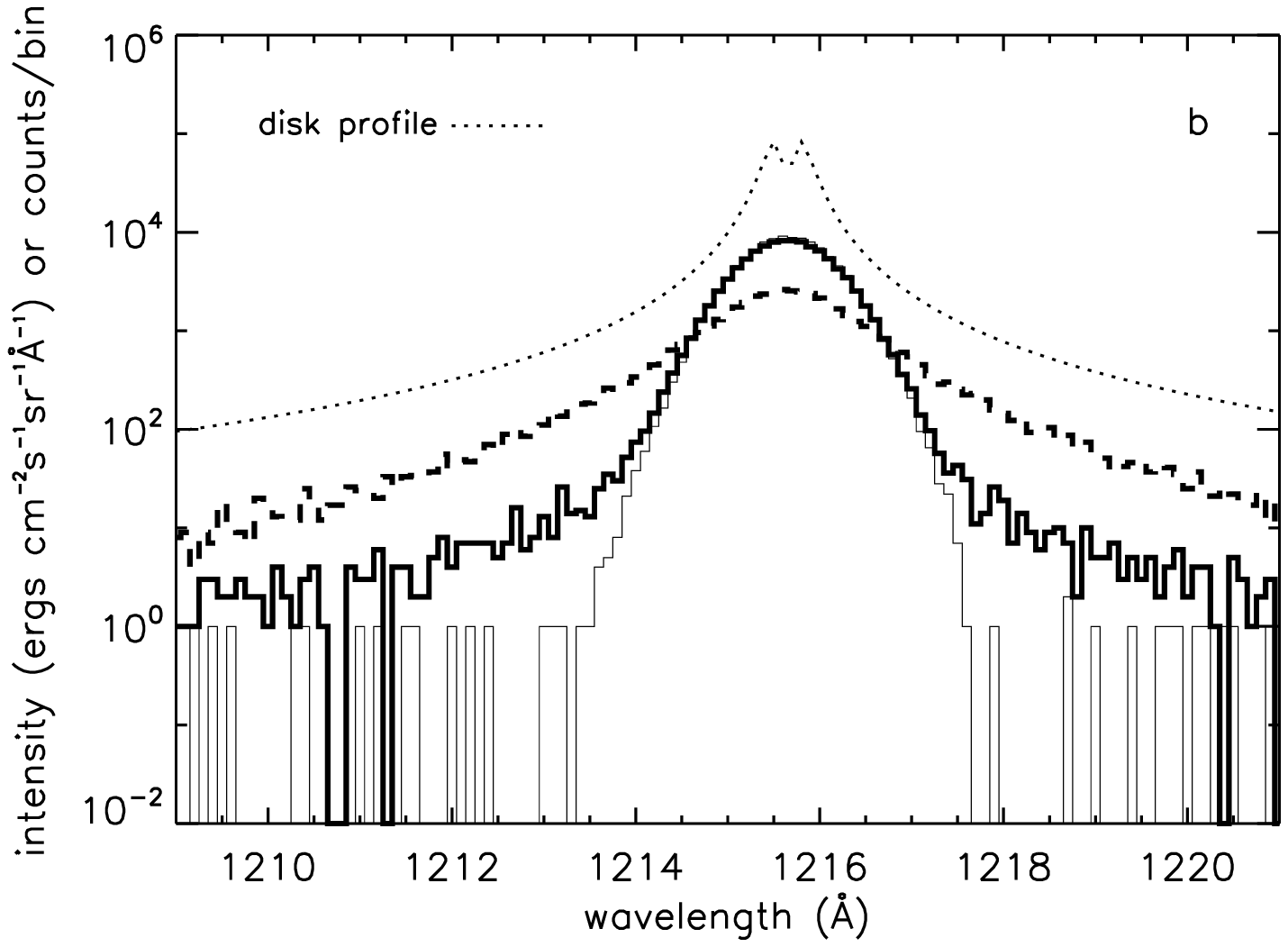}}
\vskip 20pt
\centerline{\includegraphics[scale=0.5]{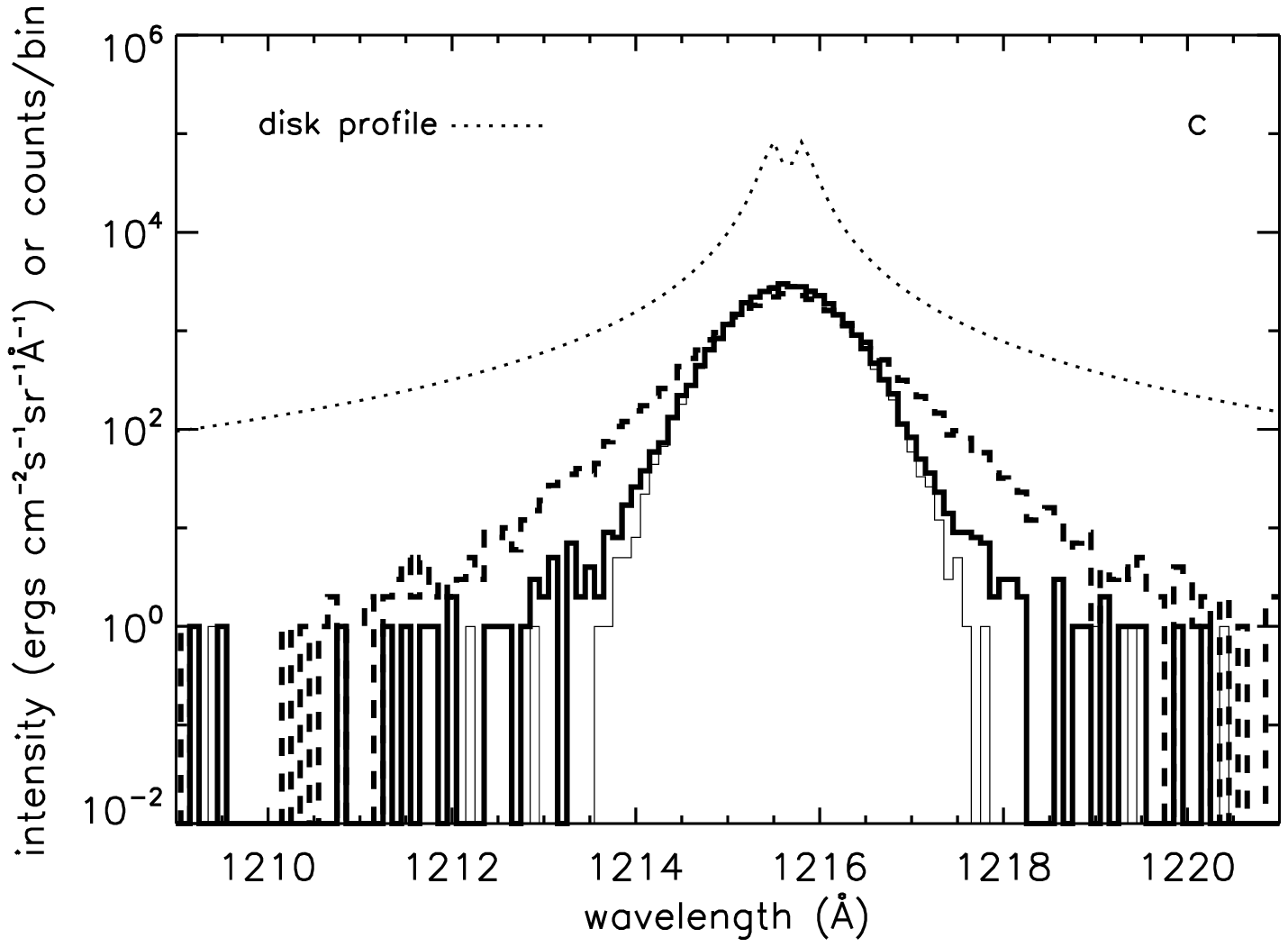}\includegraphics[scale=0.5]{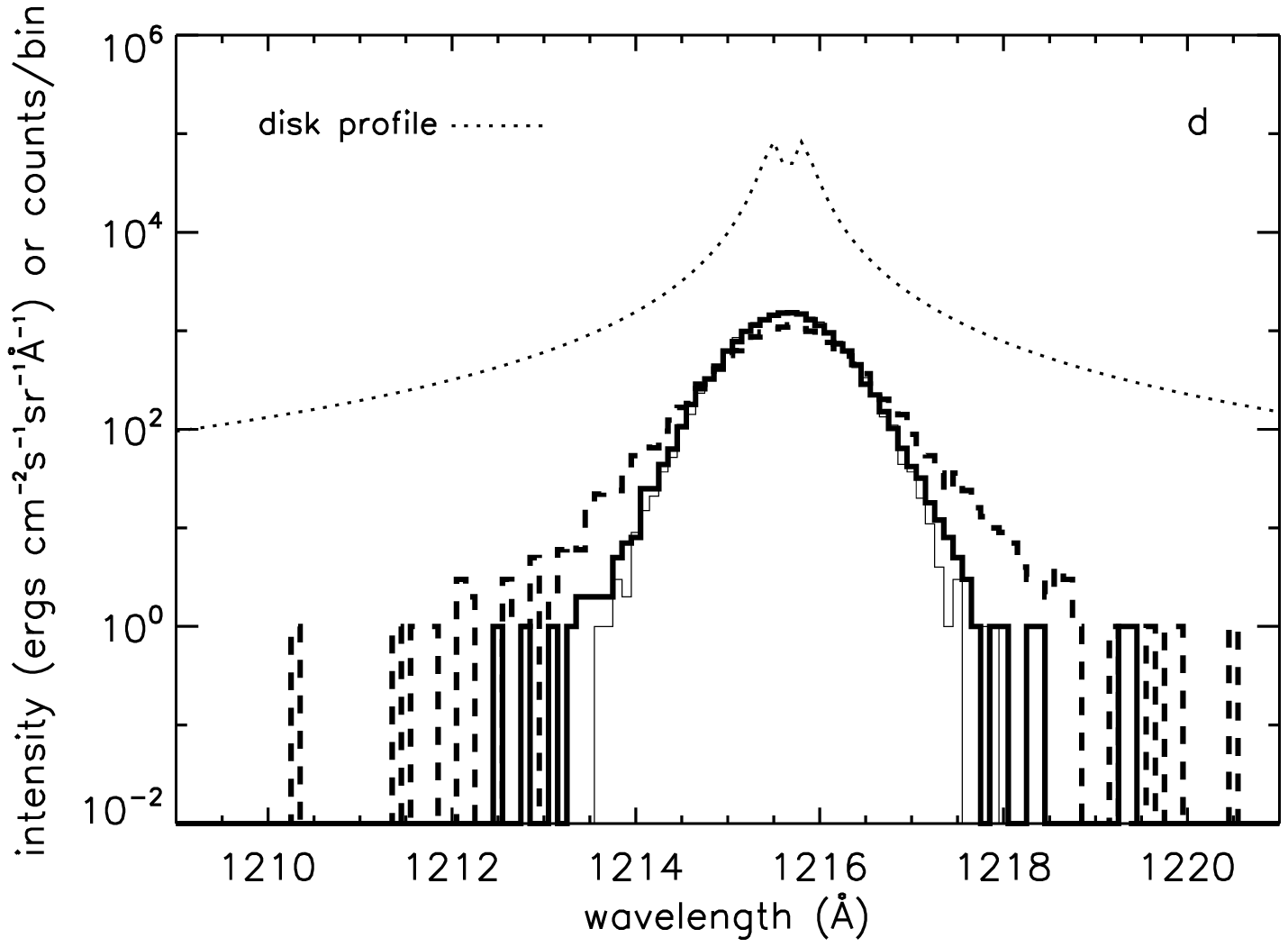}}
\caption{Simulated Lyman-$\alpha$ profiles from Figure 4a integrated for 10 and 100 s (panels
a and b), and from Figure 4b and 4c integrated for 100 s (panels c and d). The extended wings
are still detectable at these shorter integration times.
\label{fig5}}
\end{figure}

\begin{figure}
\centerline{\includegraphics[scale=0.5]{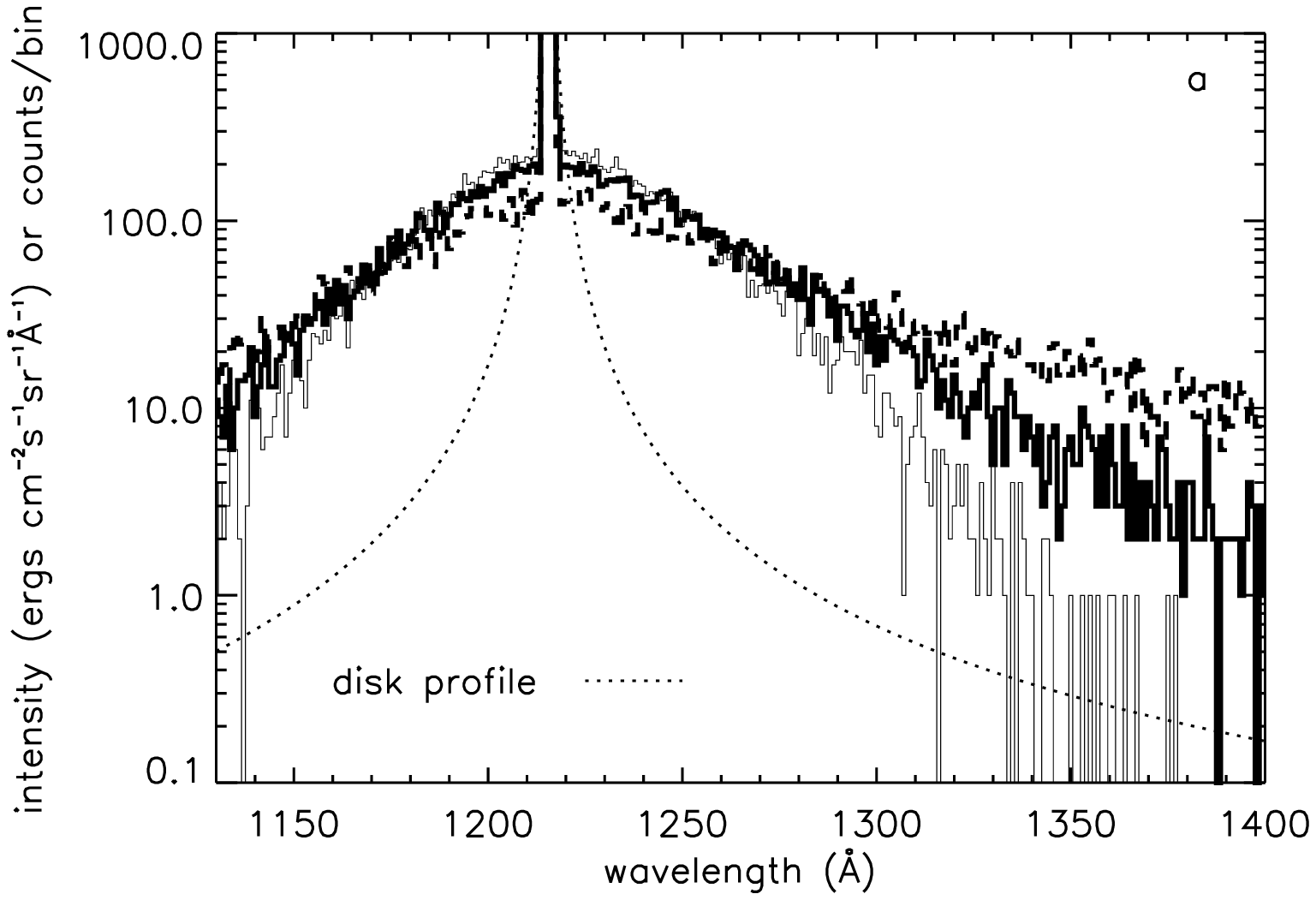}\includegraphics[scale=0.5]{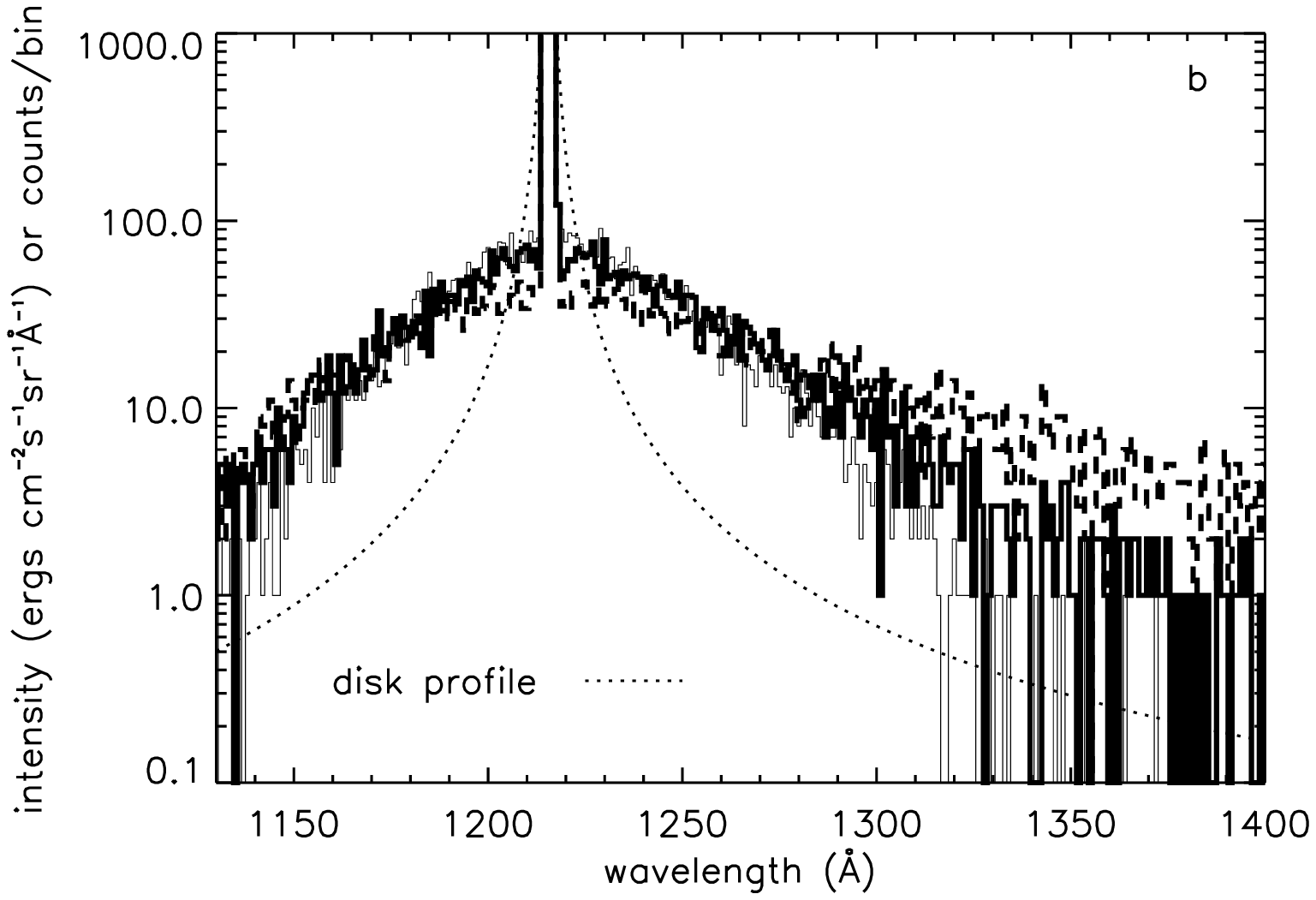}}
\vskip 20pt
\centerline{\includegraphics[scale=0.5]{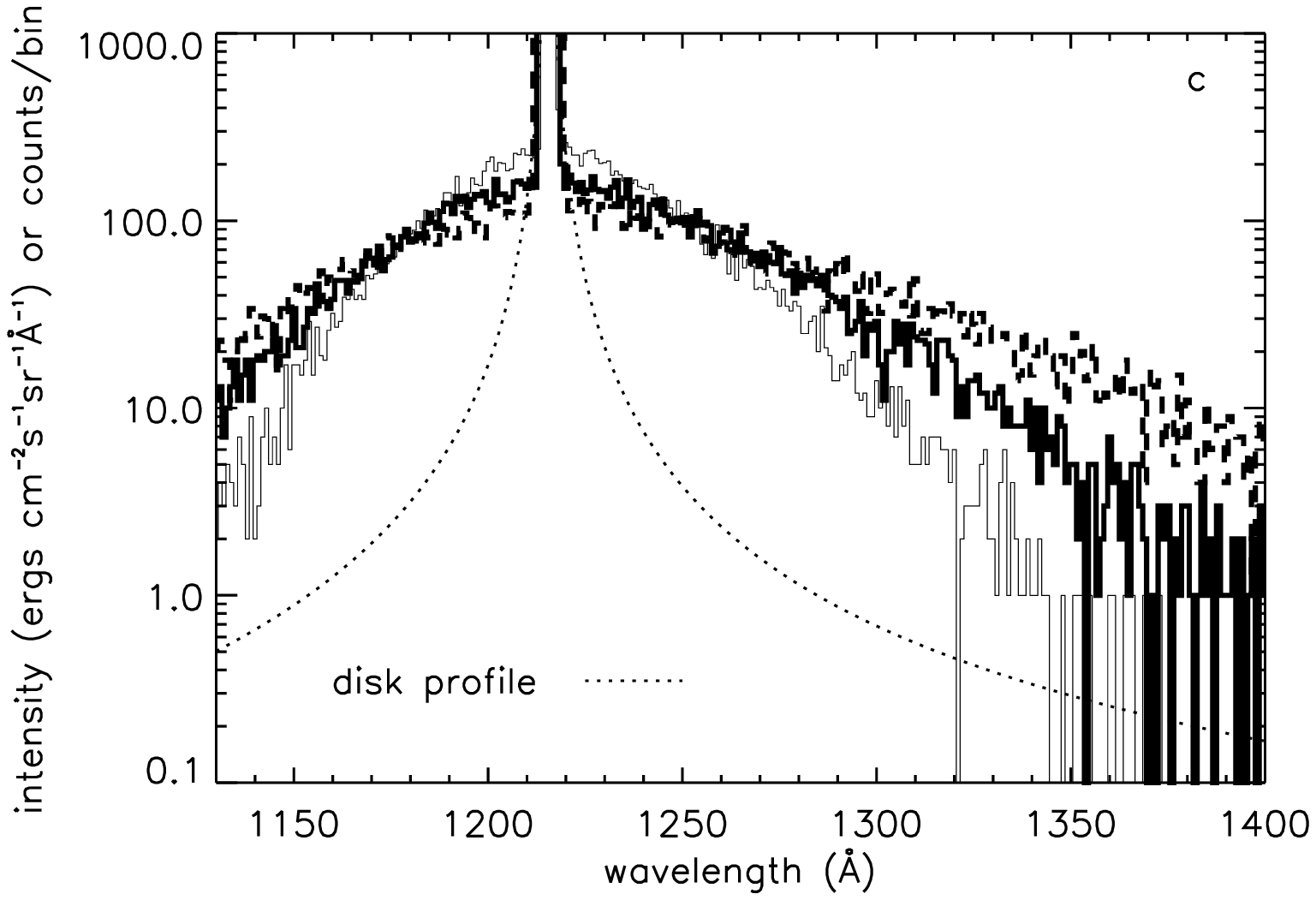}\includegraphics[scale=0.5]{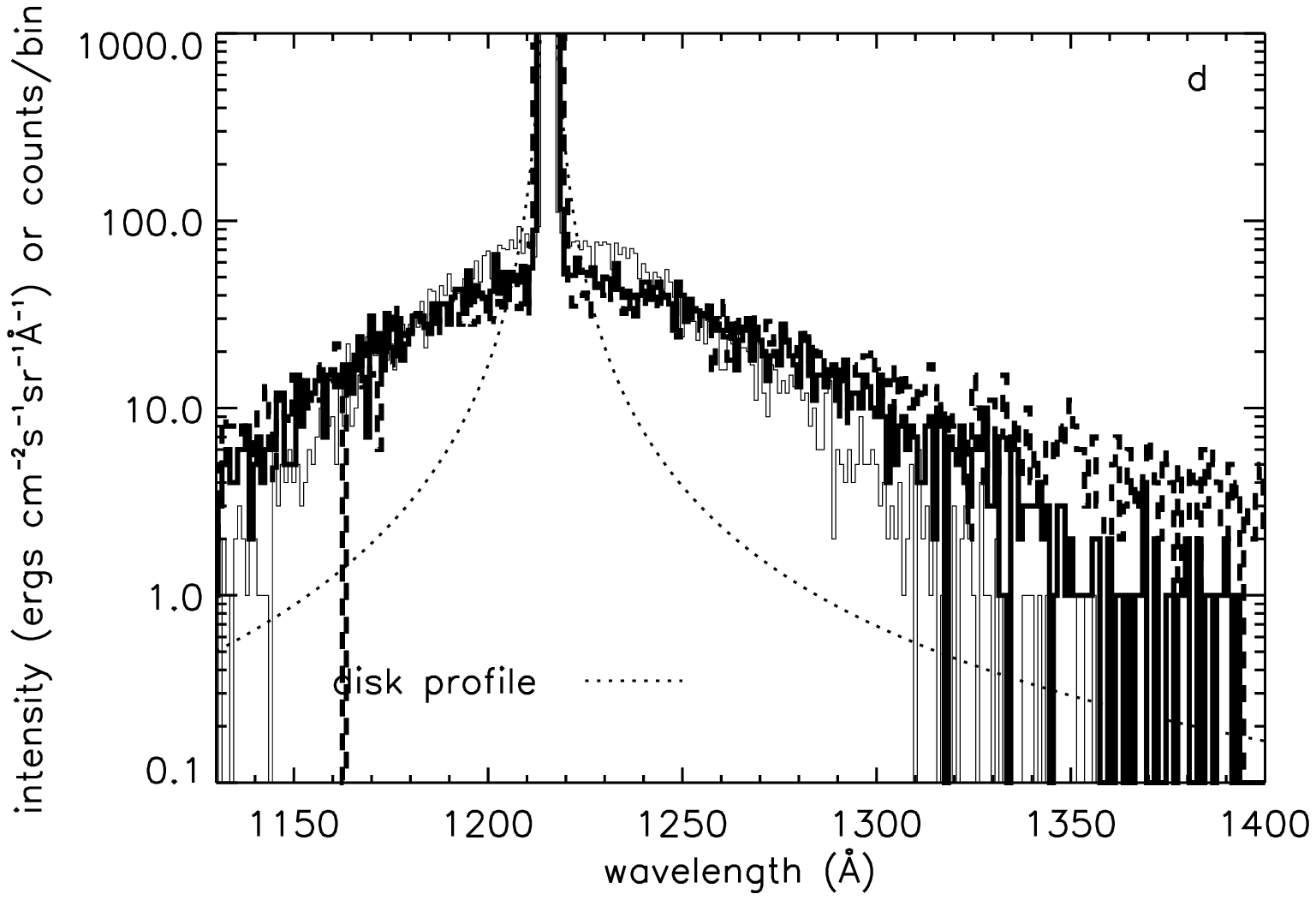}}
\caption{Simulated extended Lyman-$\alpha$ profiles showing the
effect of Thomson scattering by hot electrons. Panels a and b show
the results at 1.8 $R_{\sun}$ and 2.2 $R_{\sun}$ at $T=2\times 10^6$
K (173 eV), for a Maxwellian (thin solid histogram), $\kappa =
5$ (thick solid histogram) and $\kappa =2.4$ (thick dashed
histogram).  Panels c and d show results at 1.8 $R_{\sun}$ and
2.2 $R_{\sun}$ at $T=2\times 10^6$ K (thin solid histogram),
$T=4\times 10^6$ K (350 eV; thick solid histogram), and
$T=8\times 10^6$ K (700 eV; thick dashed histogram). The
assumed integration times is $10^5$ s, in a pixel of size $0.1
\times 1$ arcmin$^2$, determined by the angular width of a
reconnection current sheet. The histograms have 1 \AA\ bins.
\label{fig6}}
\end{figure}

\end{document}